\documentclass[prb,twocolumn,showpacs]{revtex4}
\usepackage{graphicx}
\usepackage{amsfonts}

\newcommand{\ba}{\begin{eqnarray}}
\newcommand{\be}{\begin{equation}}
\newcommand{\ea}{\end{eqnarray}}
\newcommand{\ee}{\end{equation}}
\newcommand{\ds}{\displaystyle}

\newcommand{\tr}{\mathrm{tr}}
\newcommand{\Tr}{\mathrm{Tr}\,}
\newcommand{\Lv}{\mathcal{L}}
\renewcommand{\P}{\mathcal{P}}
\newcommand{\Q}{\mathcal{Q}}
\newcommand{\rrangle}{\rangle\!\rangle}
\newcommand{\llangle}{\langle\!\langle}
\newcommand{\tP}{\tilde\mathcal{P}}
\newcommand{\tQ}{\tilde\mathcal{Q}}
\newcommand{\Lh}{\mathcal{L}_\mathrm{hyb}}
\newcommand{\LN}{\mathcal{L}_0}

\newcommand{\ignore}[1]{}

\begin{document}

\title{Time-convolutionless master equation for quantum dots:\\ Perturbative expansion to
arbitrary order}

\author{Carsten Timm}
\email{carsten.timm@tu-dresden.de}
\affiliation{Institute for Theoretical Physics, Technische Universit\"at
Dresden, 01062 Dresden, Germany}

\date{\today}

\begin{abstract}
The master equation describing the non-equilibrium dynamics of a quantum dot
coupled to metallic leads is considered. Employing a superoperator approach, we
derive an exact time-convolutionless master equation for the probabilities of
dot states, i.e., a time-convolutionless Pauli master equation. The generator of
this master equation is derived order by order in the hybridization between dot
and leads. Although the generator turns out to be closely related to the
\textit{T}-matrix expressions for the transition rates, which are plagued by
divergences, in the time-convolutionless generator all divergences cancel order
by order. The time-convolutionless and \textit{T}-matrix master equations are
contrasted to the Nakajima-Zwanzig version. The absence of divergences
in the Nakajima-Zwanzig master equation due to the nonexistence of secular
reducible contributions becomes rather transparent in our approach, which
explicitly projects out these contributions. We also show that the
time-convolutionless generator contains the generator of the Nakajima-Zwanzig
master equation in the Markov approximation plus corrections, which we make
explicit. Furthermore, it is shown that the stationary solutions of the
time-convolutionless and the
Nakajima-Zwanzig master equations are identical. However, this identity
neither extends to perturbative expansions truncated at finite order nor to
dynamical solutions. We discuss the conditions under which the
Nakajima-Zwanzig-Markov master equation nevertheless yields good results.
\end{abstract}

\pacs{
03.65.Yz, 
05.60.Gg, 
73.23.Hk, 
73.63.-b  
}

\maketitle

\section{Introduction}
\label{sec.intro}

Electronic transport through small quantum systems, such as quantum dots or single
molecules, has been intensively studied in recent years.\cite{BoW08,OBL08,AMS10}
Apart from envisioned applications, such devices address fundamental questions of
non-equilibrium quantum statistics. Quantum dots coupled to electronic leads under a
bias voltage generically relax towards a stationary state. Unless the number of
relevant degrees of freedom of the quantum dot is very small, the relaxational
dynamics is complex, including broadly distributed time scales and damped oscillatory
behavior. The stationary state that is eventually approached typically depends on the
physical parameters in a complicated way and can in particular be very different from
the equilibrium state of the isolated dot.

The descriptions of transport through quantum dots or molecules far from equilibrium have so
far followed three broad approaches. In the first, the focus is on an electron tunneling
through the device. Its dynamics is described by a non-equilibrium Green function (NEGF).
The current through the dot can be expressed in terms of the local NEGF on the dot, which
contains selfenergies due to the tunneling or hybridization between dot and
leads.\cite{MeW92} This hybridization, which is described by a bilinear component
$H_\mathrm{hyb}$ of the Hamiltonian, in principle can be incorporated
exactly. On the other hand,
interactions with other electrons, with vibrational modes, or with local spins,
which all are particularly important for small dots or single molecules, require
approximations.\cite{WiM94,HKH98,ScK00,RKW01,MAM04,PLN05,GPM07,MiM07,KRS07,%
JMS07,ElT10}

The second approach revolves around the non-equilibrium Keldysh generating
function.\cite{WET08,SMR10} It is most naturally expressed as a functional
integral and, with suitable source terms, contains the full information on
the system. This formulation is particularly suitable for numerical
calculations.
When errors due to Trotter discretization and a cutoff time for the memory
kernel are properly controlled, the results are numerically exact.

The third approach focuses on the dynamics of the small system. An
equation of motion for the reduced density operator in the Fock sub-space of the small
system is derived by integrating out the lead degrees of freedom. The result is a
\emph{master equation}
(ME).\cite{ScS94,KSS95,KSS96,BrF03,BrF04,MAM04,Tim08,LeW08,Sch09,KGL10}
If the small system is sufficiently simple, the interactions
within this system can be treated exactly. However, integrating out the lead states
naturally leads to a perturbative series in the hybridization $H_\mathrm{hyb}$.

Master equations can be either non-local or local in time. A non-local ME, for example of
Nakajima-Zwanzig (NZ) type,\cite{Nak58,Zwa60} contains a memory kernel, which relates the
rate of change of the reduced density operator at a time $t$ to the reduced density
operator at all previous times $t'<t$. On the other hand, a local
(``time-convolutionless,'' TCL) ME\cite{ToM76,STH77,Ahn94} expresses the rate of
change of the reduced density operator at time $t$ in terms of the reduced
density operator at time $t$ only.

If one has a practical method for generating all terms in the perturbation series
for the transition rates or memory kernel in orders of $H_\mathrm{hyb}$, one can hope
to resum the series or at least a subseries. This idea has been very fruitful for
many-particle physics, from the Dyson equation to the theory of the Kondo effect. For
the non-local ME of NZ-type, Schoeller, Sch\"on, and K\"onig have developed a
real-time diagrammatic scheme that generates all
terms.\cite{ScS94,KSS95,KSS96,Sch09} For
a large class of systems including a quite
general coupling Hamiltonian $H_\mathrm{hyb}$, Schoeller\cite{Sch09} has
presented a compact
superoperator formulation in Laplace space. This formulation is particularly suitable
for a non-equilibrium renormalization-group approach, which in principle includes all
orders in $H_\mathrm{hyb}$.\cite{Sch09}

Apart from the NZ ME, the \textit{T}-matrix approach from
time-dependent perturbation theory has been used to calculate the transition rates in the
ME.\cite{Ake99,BrF04,GoL04,KOO04,JoS06,ElT07,LKO08} It has the advantage
of being relatively straightforward but is known to produce divergences beyond second order
in $H_\mathrm{hyb}$, the nature of which has recently been
clarified.\cite{Tim08,BKG10,KGL10} The superoperator derivation of the
\textit{T}-matrix ME will make their origin transparent.

The TCL ME has the obvious advantage of being an \emph{exact} ME describing the full
dynamics that is nevertheless local in time. However, so far a method for
generating all terms in the perturbation series for the TCL ME has been lacking, which
has limited its usefulness.

The main purpose of the present paper is to derive an iterative scheme for constructing
all orders in the perturbative expansion of the generator of the TCL ME. The results are
valid for the exact ME describing the full dynamics. Only at the end we will
discuss the implications for the stationary state. Furthermore, a surprising
connection between the TCL generator and the \textit{T}-matrix transition rates
is uncovered. This connection introduces the divergences of the
\textit{T}-matrix rates into the expansion terms of the
TCL generator. We will show that these divergences cancel order by order.
In the present paper, we concentrate on master equations for the diagonal components
of the reduced density matrix, i.e., for the probabilities. We will call these
the \emph{Pauli} master equations or rate equations.

In the remainder of this paper, the theoretical development is presented in
Sec.\
\ref{sec.theory}. After a brief review of the superoperator formalism and the TCL ME, we
derive the Pauli version thereof, i.e., the TCL rate equations. Then we derive the
\textit{T}-matrix formula for the transition rates within the same formalism and exhibit 
the relation between the rates derived within the two approaches. After that, we
organize the perturbative expansion of the rates derived within both approaches
in a way that makes all divergent terms explicit and then show that all the
divergences cancel in the TCL rates. Finally, additional insight is gained by a
discussion of the Nakajima-Zwanzig ME and of the stationary state. The results
are summarized in Sec.\ \ref{sec.conc}, where we also draw a number of
conclusions. Several proofs are relegated to appendices.

\section{Theory}
\label{sec.theory}

\subsection{Superoperators and the TCL master equation}
\label{sub.super}

Since we will make extensive use of the superoperator formalism, we
briefly review the superoperator derivation of the TCL
ME.\cite{ToM76,STH77,Ahn94,BrP02,Tim08} To make contact with the
\textit{T}-matrix approach
and to allow for the analysis of divergences, we consider a time-dependent hybridization
between dot and leads. The Hamiltonian reads
\be
H(t) = H_0 + H_\mathrm{hyb}\,e^{\eta t} ,
\label{A2.H.2}
\ee
where $\eta$ is small and positive. $H_0=H_\mathrm{dot} + H_\mathrm{leads}$ describes
the decoupled dot and leads. For convenience we assume that the
eigenstates $|m)$ of $H_\mathrm{dot}$ are non-degenerate.\cite{endnote.degen}
As usual, the leads are represented by non-interacting Fermi seas.

The density operator $\rho$ of the full system satisfies the von Neumann
equation
\be
\dot\rho = -i[H(t),\rho] =: -i\mathcal{L}(t)\,\rho ,
\label{A2.vN.2}
\ee
where we have defined the Liouvillian $\Lv$. The resulting unitary time evolution of
$\rho$ can be expressed as
\be
\rho(t) = T_\leftarrow \exp\left(-i \int_{t_0}^t dt'\, \mathcal{L}(t')\right)\,
\rho(t_0) ,
\label{A2.rforw.2}
\ee
where $T_\leftarrow$ is the time-ordering directive.

Projection superoperators $\P$ and $\Q$ are defined by
\be
\P\rho(t) := [\mathrm{tr}_\mathrm{leads}\,\rho(t)]\otimes
  \rho_\mathrm{leads}^0
\ee
and $\Q := 1-\P$. Here, $\rho_\mathrm{leads}^0$ describes the leads in
generally separate equilibrium---each lead is in equilibrium at its own
chemical potential and possibly temperature.
We write $\mathcal{L}(t) = \LN + \Lh e^{\eta t}$ with obvious definitions and
note the identities\cite{Tim08}
\ba
\P\LN & = & \LN\P ,
\label{A2.Id.1} \\
\P\Lh\P & = & 0 .
\label{A2.Id.2}
\ea
We assume that the system was in a product state at time $t_0$ with the leads in
equilibrium, i.e., $\Q\rho(t_0)=0$. Dropping this assumption would lead
to additional terms describing the propagation of $\Q\rho(t_0)$ from time
$t_0$ to $t$. Since we are interested in the case
$t_0\to-\infty$, we do not expect the initial state to be relevant. We then obtain
\be
\mathcal{P}\rho(t) = \mathcal{P}\, T_\leftarrow \exp\left(-i \int_{t_0}^t dt'\,
\big[\LN + \Lh\,e^{\eta t'}\big]\right)\,
\mathcal{P}\rho(t_0) .
\label{A2.Pprop.2}
\ee
The time-ordered exponential is expanded as
\ba
\mathcal{P}\rho(t) & = & \mathcal{P} \sum_{\nu=0}^\infty (-i)^\nu \int_{t_0}^t
  dt_1 \int_{t_0}^{t_1} dt_2
  \cdots \int_{t_0}^{t_{\nu-1}} dt_\nu  \nonumber \\
&& {}\times e^{-i\LN\,(t-t_1)}
  \Lh e^{\eta t_1} e^{-i\LN\,(t_1-t_2)}
  \Lh e^{\eta t_2} \cdots \nonumber \\
&& {}\times \Lh e^{\eta
  t_\nu} e^{-i\LN\,(t_\nu-t_0)} \mathcal{P}\rho(t_0) .
\label{A2.Texp.2}
\ea
Equations (\ref{A2.Pprop.2}) and (\ref{A2.Texp.2}) simply represent the time
evolution of the density operator, projected with $\P$ at time $t$.

The TCL ME is derived by splitting the von Neumann equation
(\ref{A2.vN.2}) into two parts,
\ba
d\mathcal{P}\rho(t)/dt & = &
  -i\mathcal{P}\mathcal{L}(t)\mathcal{P}\rho(t)
  -i\mathcal{P}\mathcal{L}(t)\mathcal{Q}\rho(t) ,
\label{A2.vN.2a} \\
d\mathcal{Q}\rho(t)/dt & = &
  -i\mathcal{Q}\mathcal{L}(t)\mathcal{P}\rho(t)
  -i\mathcal{Q}\mathcal{L}(t)\mathcal{Q}\rho(t) .
\ea
The second equation is solved by
\ba
\lefteqn{ \mathcal{Q}\rho(t) = T_\leftarrow
  \exp\left(-i\mathcal{Q}\int_{t_0}^t dt' \,
  \mathcal{L}(t')\right)\mathcal{Q}\rho(t_0) } \nonumber \\
&& {}- i \int_{t_0}^t dt'\, T_\leftarrow \exp\left(-i\mathcal{Q}\int_{t'}^t dt'' \,
  \mathcal{L}(t'')\right) \mathcal{Q}\mathcal{L}(t')\mathcal{P} \rho(t') ,
  \nonumber \\[-1ex]
&&
\label{A2.sol2.2}
\ea
where the first term vanishes under our assumption of $\Q\rho(t_0)=0$.
The main idea of the TCL approach\cite{ToM76,STH77} is to express $\rho(t')$
by propagating the full density operator backward in time,
\be
\rho(t') = T_\rightarrow \exp\left(i \int_{t'}^t dt'' \,
  \mathcal{L}(t'')\right)\, \rho(t) ,
\label{A2.rhoback.2}
\ee
where $T_\rightarrow$ is the anti-time-ordering directive. Insertion
into Eq.\ (\ref{A2.sol2.2}) gives
\ba
\lefteqn{ \mathcal{Q}\rho(t) =
  - i \int_{t_0}^t dt'\, T_\leftarrow \exp\left(-i\mathcal{Q}\int_{t'}^t dt'' \,
  \mathcal{L}(t'')\right) \mathcal{Q}\mathcal{L}(t')\mathcal{P} } \nonumber \\
&& {}\times T_\rightarrow \exp\left(i \int_{t'}^t dt'' \,
  \mathcal{L}(t'')\right)\, [\mathcal{P}\rho(t) + \mathcal{Q}\rho(t)] .
  \hspace{3.5em}
\ea
Solving for $\mathcal{Q}\rho(t)$ yields
\be
\mathcal{Q}\rho(t) = [1-\Sigma(t,t_0)]^{-1} \Sigma(t,t_0)\, \mathcal{P}\rho(t)
\label{A2.Qrho.5}
\ee
with
\ba
\lefteqn{ \Sigma(t,t_0) := -i \int_{t_0}^t dt'\, T_\leftarrow
\exp\left(-i\mathcal{Q}\int_{t'}^t dt'' \,
  \mathcal{L}(t'')\right) } \nonumber \\
&& {}\times \mathcal{Q}\mathcal{L}(t')\mathcal{P}\,
  T_\rightarrow \exp\left(i \int_{t'}^t dt'' \,
  \mathcal{L}(t'')\right) . \hspace{5em}
\ea
Inserting $\mathcal{Q}\rho(t)$ into Eq.\ (\ref{A2.vN.2a}) results in an
equation of motion for $\mathcal{P}\rho$ alone,
\be
d\mathcal{P}\rho(t)/dt =
-i\mathcal{P}\mathcal{L}(t)\,[1-\Sigma(t,t_0)]^{-1}\, \mathcal{P}\rho(t) .
\ee
This is the TCL ME for the case $\mathcal{Q}\rho(t_0)=0$.
Using Eqs.\ (\ref{A2.Id.1}) and (\ref{A2.Id.2}) and noting that
$\Sigma(t,t_0)$ contains a projection $\mathcal{Q}$ to the left, we can
rewrite this more specifically as
\ba
\lefteqn{ d\mathcal{P}\rho(t)/dt
  = -i\mathcal{P} \LN \mathcal{P}\rho(t) } \nonumber \\
&& {}-i\mathcal{P}\Lh e^{\eta t}\,[1-\Sigma(t,t_0)]^{-1}\,
  \mathcal{P}\rho(t)
\label{A2.TCLME.3}
\ea
with
\ba
\lefteqn{ \Sigma(t,t_0) = -i \mathcal{Q} \int_{t_0}^t dt' }
  \nonumber \\
&& {}\times T_\leftarrow
  \exp\left(-i \int_{t'}^t dt'' \, \big[\LN + \mathcal{Q}
  \Lh e^{\eta t''}\mathcal{Q} \big]\right)
  \nonumber \\
&& {}\times \Lh e^{\eta t'} \mathcal{P}\,
  T_\rightarrow \exp\left(i \int_{t'}^t dt'' \,
  \big[\LN + \Lh e^{\eta t''}\big]\right) . \hspace{1.5em}
\ea
The first term in Eq.\ (\ref{A2.TCLME.3}) describes the unperturbed time
evolution, while the second stems from the hybridization.

\subsection{The TCL Pauli master equation}
\label{sub.TCL}

We here derive an exact TCL ME for the diagonal components of
the reduced density operator. Since we are assuming non-degenerate dot states,
this is equivalent to an equation for the secular part of the reduced density
operator.\cite{KGL10} A ME for the diagonal
components of the density operator, albeit non-local in time, has been
derived by Zwanzig\cite{Zwa60} and rediscovered, in the context of transport,
by Leijnse and Wegewijs.\cite{LeW08}

We introduce new projection operators by
\be
\tP\rho(t) := \bigg[\sum_m |m)(m| \mathrm{tr}_\mathrm{leads}\,\rho(t)
  |m)(m|\bigg]\otimes \rho_\mathrm{leads}^0 ,
\ee
where the $|m)$ are the unperturbed dot eigenstates,
and $\tQ := 1-\tP$. $\tP$ evidently projects the density operator onto a product
form with \emph{diagonal} reduced density operator.\cite{Zwa60}
We will call $\tP\rho$ the \emph{diagonally projected} density operator.
It is easy to show that
\be
\tP\LN = \LN\tP = 0 ,
\label{A4.Id.1a}
\ee
which goes beyond Eq.\ (\ref{A2.Id.1}) for $\P$. Since $H_\mathrm{hyb}$
changes the electron number in the leads by $\pm 1$, we have
\be
\tP\Lh\tP = 0 .
\label{A4.Id.2}
\ee
We now assume that the system was in a product state at time $t_0$ with the
leads in generally separate equilibrium and \emph{diagonal} reduced density
operator, $\tQ\rho(t_0)=0$.

Repeating the derivation in Sec.\ \ref{sub.super} with the new projections
$\tP$, $\tQ$, we obtain
\be
d\tP\rho(t)/dt
  = -i\tP\Lh e^{\eta t}\,[1-\tilde\Sigma(t,t_0)]^{-1}\,
  \tP\rho(t)
\label{A4.TCLME.2}
\ee
with
\ba
\lefteqn{ \tilde\Sigma(t,t_0) := -i \tQ \int_{t_0}^t dt' }
  \nonumber \\
&& {}\times T_\leftarrow
  \exp\left(-i \int_{t'}^t dt'' \, \big[\LN + \tQ
  \Lh e^{\eta t''} \tQ \big]\right)   \nonumber \\
&& {}\times \Lh e^{\eta t'} \tP\, T_\rightarrow \exp\left(i \int_{t'}^t dt'' \,
  \big[\LN + \Lh e^{\eta t''}\big]\right) . \hspace{1.5em}
\ea
Due to Eq.\ (\ref{A4.Id.1a}), the unperturbed time evolution has dropped out of
Eq.\ (\ref{A4.TCLME.2}). We can now write
\be
d\tP\rho(t)/dt = \tilde\mathcal{S}(t,t_0)\, \tP\rho(t)
\label{A4.TCLME.4}
\ee
with the generator
\be
\tilde\mathcal{S}(t,t_0) = -i\tP\Lh e^{\eta t}\,
  [1-\tilde\Sigma(t,t_0)]^{-1} \tP .
\label{A4.tS.2}
\ee
Equation (\ref{A4.TCLME.4}) is an exact ME for the diagonally projected density operator
under the condition $\tQ\rho(t_0)=0$. It is evidently local in time. Since it
only involves the diagonal components, we call it the TCL \emph{Pauli} ME.
A Pauli ME in the reduced Fock space of the dot is of course obtained by
taking the trace over the lead degrees of freedom,
\be
d\rho_\mathrm{dot}/dt = \mathrm{tr}_\mathrm{leads}
  \tilde\mathcal{S}(t,t_0)\, \rho_\mathrm{dot}(t) \otimes \rho^0_\mathrm{leads}
  =: \tilde\mathcal{S}_\mathrm{dot}\, \rho_\mathrm{dot}(t) .
\ee
The reduced generator $\tilde\mathcal{S}_\mathrm{dot}$ written in the dot eigenbasis
is the transition-rate matrix. Ensembles of such matrices
are studied in Ref.\ \onlinecite{Tim09} within random matrix theory.

We have now eliminated the off-diagonal components of the reduced density matrix
$\rho_\mathrm{dot}$ from the equations of motion, similar in spirit to Zwanzig's
work\cite{Zwa60} and also to Refs.\ \onlinecite{LeW08} and \onlinecite{KGL10}. We are
therefore able to determine the dynamics of the probabilities of dot states exclusively from
the knowledge of these probabilities at a given time. This does not mean that we assume
the off-diagonal components to be small, which is not generally true.

The knowledge of the probabilities is sufficient for the
calculations of dot observables that commute with the dot Hamiltonian
$H_\mathrm{dot}$. To see this, we denote the operator for such an observable by
$A$ in the Fock space of the dot. Then the operator in the Fock
space of the whole system is $A\otimes 1_\mathrm{leads}$ in an obvious
notation. The average of the observable is
\ba
\langle A\rangle(t) & = & \Tr \rho(t) A \otimes 1_\mathrm{leads}
  = \tr_\mathrm{dot}\, \rho_\mathrm{dot}(t)\, A \nonumber \\
& = & \sum_{mn} \rho^\mathrm{dot}_{mn}(t)\, A_{nm} ,
\ea
writing matrix elements of dot states $|m)$, $|n)$ as
$\rho^\mathrm{dot}_{mn} = (m|\rho_\mathrm{dot}|n)$ etc. If $A$ commutes with
the dot Hamiltonian we can choose $A$ to be diagonal
in the dot eigenbasis $\{|m)\}$ so that
\be
\langle A\rangle(t) = \sum_m \rho^\mathrm{dot}_{mm}(t)\, A_{mm}
= \mathrm{Tr}\, \tP\rho(t)\,A .
\ee
Thus the knowledge of $\tP\rho(t)$ is sufficient to calculate the average. Examples
are the charge on the dot, the vibrational energy of
a molecule, or the component of its spin parallel to an applied magnetic field,
assuming vanishing transverse anisotropy.
On the other hand, the current does not commute with
$H_\mathrm{dot}$ and thus does depend on the off-diagonal components.\cite{ScO09}
However, it is possible to reconstruct the full density operator from $\tP\rho(t)$,
\be
\rho(t) = \tP\rho(t)+\tQ\rho(t) = [1-\tilde\Sigma(t,t_0)]^{-1}
  \,\tP\rho(t) ,
\ee
compare Eq.\ (\ref{A2.Qrho.5}).

\subsection{Superoperator derivation of the \textit{T}-matrix formula}
\label{sub.Tm}

In the following, the \textit{T}-matrix formula for the transition rates is rederived
within the superoperator formalism to allow a direct comparison with
the exact TCL ME. Moreover, we show that this derivation relies on a single
straightforward but generally unjustified approximation.

To start with, note that the exponential time
dependence of the perturbation in Eq.\ (\ref{A2.H.2}) is exactly the case considered
by Bruus and Flensberg\cite{BrF04} in their derivation of the \textit{T}-matrix
formula. The analog of Eqs.\ (\ref{A2.Pprop.2}) and (\ref{A2.Texp.2}) for diagonal
projection reads
\ba
\lefteqn{ \tP\rho(t) = \tP\, T_\leftarrow \exp\left(-i \int_{t_0}^t dt'\,
  \big[\LN + \Lh\,e^{\eta t'}\big]\right)\,
  \tP\rho(t_0) } \nonumber \\
&& = \tP \sum_{\nu=0}^\infty (-i)^\nu \int_{t_0}^t dt_1
  \int_{t_0}^{t_1} dt_2
  \cdots \int_{t_0}^{t_{\nu-1}} dt_\nu \,
  \Lh e^{\eta t_1} \nonumber \\
&& {}\times e^{-i\LN\,(t_1-t_2)}
  \Lh e^{\eta t_2} \cdots
  \Lh e^{\eta t_\nu} \tP\rho(t_0) ,
\label{A6.Texp.2}
\ea
where we have used Eq.\ (\ref{A4.Id.1a}). This is the time evolution of the full
density operator under the condition $\tQ\rho(t_0)=0$, projected with $\tP$ at time $t$.

Taking the time derivative of Eq.\ (\ref{A6.Texp.2}), we obtain
\be
d\tP\rho(t)/dt = \tilde\mathcal{R}(t,t_0)\, \tP\rho(t_0)
\label{A6.dtPr.ex.3}
\ee
with
\ba
\lefteqn{ \tilde\mathcal{R}(t,t_0) := -i\tP \Lh e^{\eta t}
  \sum_{\mu=0}^\infty (-i)^\mu
  \int_{t_0}^t dt_1 \int_{t_0}^{t_1} dt_2
  \cdots } \nonumber \\
&& {}\times \int_{t_0}^{t_{\mu-1}} dt_\mu\,
  e^{-i\LN\,(t-t_1)} \Lh e^{\eta t_1}
  e^{-i\LN\,(t_1-t_2)} \nonumber \\
&& {}\times \Lh e^{\eta t_2}
  \cdots \Lh e^{\eta t_\mu} \tP ,
\label{A4.Rfull.3}
\ea
where the new summation index is $\mu=\nu-1$ and the integration variables are
now counted by $\mu$. The terms in this series are of order $\mu+1$ in $\Lh$.
In the limit $t_0\to -\infty$, it is straightforward to evaluate the time
integrals at non-zero $\eta$,
\ba
\lefteqn{ \tilde\mathcal{R}(t,-\infty) = -i \sum_{\nu=1}^\infty e^{\nu\eta t}\,
  \tP \Lh\, [-\LN + (\nu-1)i\eta]^{-1}\, \Lh } \nonumber \\
&& {}\times [-\LN + (\nu-2)i\eta]^{-1}\, \Lh \cdots \nonumber \\
&& {}\times \Lh\,(-\LN+i\eta)^{-1}\, \Lh\tP .\hspace{10.5em}
\label{A4.Rfull.4}
\ea
The terms are of order $\nu$ in $\Lh$. Due to the trace over lead states coming
from the leftmost $\tP$ and the equilibrium lead density operator
$\rho_\mathrm{leads}^0$ contained in the rightmost $\tP$, this
expression contains \emph{equilibrium} averages of products of lead electron
creation operators $a^\dagger_{\alpha\mathbf{k}\sigma}$ and annihilation operators
$a_{\alpha\mathbf{k}\sigma}$. To obtain a non-zero contribution, these
operators must be paired. Since the hybridization
Hamiltonian $H_\mathrm{hyb}$ is linear in these operators, only terms of even
order $\nu$ contribute to $\tilde\mathcal{R}(t,-\infty)$.

We will now show that the central approximation of the \textit{T}-matrix
approach consists of taking $\tilde\mathcal{R}(t,t_0)$ to be the generator of a
Pauli ME that is local in time,
\be
d\tP\rho(t)/dt \approx \tilde\mathcal{R}(t,t_0)\, \tP\rho(t) .
\label{A6.dtPr.T.3}
\ee
To that end, we show that this ME indeed leads to the usual \textit{T}-matrix
formula.\cite{BrF04}

Expressing the Liouvillians in Eq.\ (\ref{A4.Rfull.3}) in terms of the
corresponding Hamiltonians, we obtain
\ba
\lefteqn{ \tilde\mathcal{R}(t,t_0)\bullet = -i
  \tP\, \bigg\{ \sum_{\mu=0}^\infty
  (-i)^\mu \int_{t_0}^t dt_1 \int_{t_0}^{t_1} dt_2 \cdots \int_{t_0}^{t_{\mu-1}}
  \!\! dt_\mu } \nonumber \\
&& {}\times \Big[ H_\mathrm{hyb} e^{\eta t},
  e^{-i H_0\,(t-t_1)}\, [H_\mathrm{hyb} e^{\eta t_1},
  e^{-i H_0\,(t_1-t_2)} \nonumber \\
&& {}\times [H_\mathrm{hyb} e^{\eta t_2},
  \cdots
  [H_\mathrm{hyb} e^{\eta t_\mu},e^{-i H_0\,(t_\mu-t_0)}\, {\tP\bullet}
  \nonumber \\
&& {}\times e^{i H_0\,(t_\mu-t_0)} ]
  \cdots ] e^{i H_0\,(t_1-t_2)} ]\,
  e^{i H_0\,(t-t_1)}
  \Big] \bigg\} .\hspace{4em}
\label{A4.Rrho.asH.2}
\ea
We now consider unequal initial and final eigenstates, $|i\rangle$ and $|f\rangle$,
respectively, of $H_0$. Pure initial and final
states are described by the density operators $|i\rangle\langle i|$ and
$|f\rangle\langle f|$, respectively. Expanding the nested commutators, except
for the outermost one, we obtain for the matrix element of
$\tilde\mathcal{R}(t,t_0)$ between these pure states
\ba
\lefteqn{ \Gamma_{fi} := \langle
  f|\big\{ \tilde\mathcal{R}(t,t_0)|i\rangle\langle
  i|\big\} |f\rangle } \nonumber \\
&& = -i \, \langle f| \sum_{\mu,\nu=0}^\infty
  (-i)^\mu i^\nu \int_{t_0}^t dt_1 \int_{t_0}^{t_1} dt_2 \cdots
  \int_{t_0}^{t_{\mu-1}} \!\! dt_\mu \nonumber \\
&& {}\times
  \int_{t_0}^t dt_1' \int_{t_0}^{t_1'} dt_2' \cdots
  \int_{t_0}^{t_{\nu-1}'} \!\! dt_\nu'\,
  \Big\{ H_\mathrm{hyb} e^{\eta t},
  e^{-i H_0\,(t-t_1)} \nonumber \\
&& {}\times H_\mathrm{hyb} e^{\eta t_1}
  e^{-i H_0\,(t_1-t_2)} H_\mathrm{hyb} e^{\eta t_2} \cdots
  H_\mathrm{hyb} e^{\eta t_\mu} \nonumber \\
&& {}\times e^{-i H_0\,(t_\mu-t_0)} \, |i\rangle\langle i| \,
  e^{i H_0\,(t_\nu'-t_0)}
  H_\mathrm{hyb} e^{\eta t_\nu'}
  \cdots \nonumber \\
&& {}\times H_\mathrm{hyb} e^{\eta t_2'}
  e^{i H_0\,(t_1'-t_2')} H_\mathrm{hyb} e^{\eta t_1'}
  e^{i H_0\,(t-t_1')}
  \Big\} |f\rangle .
\ea
It is helpful to rewrite this expression as a derivative,
\ba
\lefteqn{ \Gamma_{fi}
  = \frac{d}{dt}\: \bigg| \langle f| \sum_{\mu=0}^\infty
  (-i)^\mu \int_{t_0}^t dt_1 \int_{t_0}^{t_1} dt_2 \cdots
  \int_{t_0}^{t_{\mu-1}} dt_\mu } \nonumber \\
&& {}\times e^{i H_0 t_1}\, H_\mathrm{hyb} e^{\eta
  t_1} e^{-i H_0\,(t_1-t_2)} H_\mathrm{hyb} e^{\eta t_2} \cdots
  \nonumber \\
&& {}\times H_\mathrm{hyb} e^{\eta t_\mu}
  e^{-i H_0 t_\mu} \, |i\rangle \bigg|^2 .\hspace{10.5em}
\ea
Next, the initial time $t_0$ is sent to $-\infty$ at finite $\eta$.
With $\tau_1 = t-t_1$ and $\tau_\mu = t_{\mu-1} - t_\mu$ for $\mu>1$ we obtain
\ba
\lefteqn{ \Gamma_{fi} = \frac{d}{dt}\, \bigg|\langle f|\, e^{iH_0 t}
  \sum_{\mu=1}^\infty
  (-i)^\mu \int_0^\infty d\tau_1 \int_0^\infty d\tau_2 \cdots
  \int_0^\infty d\tau_\mu } \nonumber \\
&& {}\times e^{-iH_0 \tau_1} H_\mathrm{hyb} e^{-iH_0\tau_2}
  H_\mathrm{hyb} e^{-iH_0\tau_3}
  \cdots  e^{-iH_0\tau_\mu} H_\mathrm{hyb} \nonumber \\
&& {}\times e^{-iH_0 (t-\tau_1-\tau_2-\ldots-\tau_\mu)} \,
  e^{\eta [\mu t-\mu\tau_1-(\mu-1)\tau_2-\ldots-\tau_\mu]} \,
  |i\rangle \bigg|^2 \nonumber \\
&& = \frac{d}{dt}\, \bigg|  \langle f|\,
  \sum_{\mu=1}^\infty e^{\mu\eta t} \frac{1}{E_i - E_f + i\mu\eta} \,
  H_\mathrm{hyb} \nonumber \\
&& {}\times \frac{1}{E_i - H_0 + i(\mu-1)\eta} \,
  H_\mathrm{hyb} \cdots \frac{1}{E_i - H_0 + i\eta} \nonumber \\
&& {}\times H_\mathrm{hyb} \, |i\rangle \bigg|^2 .
\label{Tm.fi2.3}
\ea
We have used that the $\mu=0$ term vanishes for $|f\rangle \neq
|i\rangle$. The fractions are to be understood as inverse ordinary
operators. The time derivative can now be evaluated,
\ba
\lefteqn{ \Gamma_{fi} = \sum_{\mu=1}^\infty \sum_{\nu=1}^\infty
  \frac{(\mu+\nu)\,\eta\, e^{\mu\eta t} e^{\nu\eta t}}
  {(E_i-E_f-i\mu\eta)(E_i-E_f+i\nu\eta)} } \nonumber \\
&& {}\times  \langle i| H_\mathrm{hyb} \, \frac{1}{E_i - H_0 - i\eta}\, \cdots
  \nonumber \\
&& {}\times H_\mathrm{hyb}\,
  \frac{1}{E_i - H_0 - i(\mu-1)\eta}\, H_\mathrm{hyb} |f\rangle
  \nonumber \\
&& {}\times \langle f|\,
  H_\mathrm{hyb} \frac{1}{E_i - H_0 + i(\nu-1)\eta} \,
  H_\mathrm{hyb} \cdots \nonumber \\
&& {}\times \frac{1}{E_i - H_0 + i\eta}\,
  H_\mathrm{hyb} \, |i\rangle .\hspace{8em}
\label{Tm.Gafi.5}
\ea
We notice that the limit $\eta\to 0^+$ can be taken in the factors
$(E_i-H_0\pm i\kappa\eta)^{-1}$ independently from the first factor
under the sum.
In the former, $\eta>0$ indicates in which complex half plane the
poles are located. In the latter, the limit $\eta\to 0^+$ leads to a
$\delta$-function implementing energy conservation,
\ba
\lefteqn{ \Gamma_{fi} = \sum_{\mu=1}^\infty \sum_{\nu=1}^\infty
  2\pi\, \delta(E_i-E_f) } \nonumber \\
&& {}\times \langle i| \left( H_\mathrm{hyb} \, \frac{1}{E_i - H_0 - i0^+}
  \right)^{\mu-1} H_\mathrm{hyb}\, |f\rangle \nonumber \\
&& {}\times \langle f|\, H_\mathrm{hyb}
  \left( \frac{1}{E_i - H_0 + i0^+} \,
  H_\mathrm{hyb} \right)^{\nu-1} |i\rangle .
\ea
Since $H_\mathrm{hyb}$ changes the electron number in the leads by $\pm 1$,
$\Gamma_{fi}$ can only be non-zero if $\mu$ and $\nu$ are both even or both odd.

Defining the \textit{T}-matrix
\be
T := \sum_{\mu=1}^\infty H_\mathrm{hyb}
  \left( \frac{1}{E_i - H_0 + i0^+}\,H_\mathrm{hyb} \right)^{\mu-1} ,
\label{Tm.Tdef.1}
\ee
we obtain the well-known result\cite{BrF04}
\be
\Gamma_{fi} = 2\pi\, \delta(E_i-E_f)\, |\langle f|T|i\rangle|^2 .
\label{Tm.FGR.2}
\ee
Note that we have obtained this result explicitly for the exponential time
dependence of the hybridization. It was not necessary to consider a different
time dependence at intermediate steps, as in Ref.\ \onlinecite{BrF04}.

We now use a product basis of unperturbed eigenstates $|m)$, $|n)$ of the dot
and $|i\rrangle$, $|f\rrangle$ of the leads. Summing over all initial lead
states $|i\rrangle$ and final lead states $|f\rrangle$, we obtain
the \textit{T}-matrix expression for the transition rate from dot state $|n)$ to
dot state $|m)\neq |n)$,
\be
\tilde R_{n\to m} = 2\pi \sum_{i,f} W_i\,
  \big| \llangle f| (m| T |n) |i\rrangle \big|^2\,
  \delta(E_n + \epsilon_i - E_m - \epsilon_f) .
\label{Tm.R.app4}
\ee
Here, $E_m$ ($\epsilon_i$) are eigenenergies of dot (lead) states and
$W_i$ is the equilibrium probability to find the leads in state
$|i\rrangle$. The sums over lead states are understood
as integrals if their spectrum is continuous.

We have shown that the \textit{T}-matrix formula (\ref{Tm.R.app4}) for the
transition rates is what one gets if one takes the exact time evolution of the
density operator, projects onto diagonal density operators of product form with
the leads in equilibrium, and then \emph{by hand} replaces the projected density
operator at the initial time, $\tP\rho(t_0)$, by the projected density operator
at the present time, $\tP\rho(t)$. This confirms the statement made in Ref.\
\onlinecite{Tim08} that the \textit{T}-matrix approach to transport
misinterprets the transition rates between dot states $|n)$ at time
$t_0\to-\infty$ and $|m)$ at time $t$ as transition rates between $|n)$ and
$|m)$ both at time $t$.

\subsection{Relation between TCL Pauli and \textit{T}-matrix generators}
\label{sub.rel}

We derive two simple relations between the generators $\tilde\mathcal{S}$
and $\tilde\mathcal{R}$. The defining equations (\ref{A4.TCLME.4}) and
(\ref{A6.dtPr.ex.3}) read
\begin{eqnarray*}
d\tP\rho(t)/dt & = & \tilde\mathcal{S}(t,t_0)\,\tP\rho(t) , \\
d\tP\rho(t)/dt & = & \tilde\mathcal{R}(t,t_0)\,\tP\rho(t_0) .
\end{eqnarray*}
The first equation is solved by
\be
\tP\rho(t_1) = T_\leftarrow \exp\left(\int_{t_2}^{t_1} dt'\,
  \tilde\mathcal{S}(t',t_0)\right)\, \tP\rho(t_2) ,
\label{SR.Prhot.3}
\ee
where $t_1\ge t_2$. Choosing $t_1=t$ and $t_2=t_0$
and taking the time derivative we obtain
\be
\frac{d}{dt}\,\tP\rho(t) = \tilde\mathcal{S}(t,t_0)\, T_\leftarrow
\exp\left(\int_{t_0}^t dt'\,
  \tilde\mathcal{S}(t',t_0)\right)\, \tP\rho(t_0) .
\ee
Comparison with Eq.\ (\ref{A6.dtPr.ex.3}) yields the identity
\be
\tilde\mathcal{R}(t,t_0) = \tilde\mathcal{S}(t,t_0)\, T_\leftarrow
  \exp\left(\int_{t_0}^t dt'\, \tilde\mathcal{S}(t',t_0)\right) .
\label{SR.results.1}
\ee
Conversely, to represent $\tilde\mathcal{S}$ in terms of $\tilde\mathcal{R}$,
we integrate Eq.\ (\ref{A6.dtPr.ex.3}) from time $t_0$ to $t$,
\be
\tP\rho(t) = \tP\rho(t_0) + \int_{t_0}^t dt'\,
  \tilde\mathcal{R}(t',t_0)\,\tP\rho(t_0) .
\ee
Comparison with Eq.\ (\ref{SR.Prhot.3}) yields
\be
T_\leftarrow \exp\left(\int_{t_0}^t dt'\,
  \tilde\mathcal{S}(t',t_0)\right)
  = 1 + \int_{t_0}^t dt'\, \tilde\mathcal{R}(t',t_0) .
\label{SR.eSR.2}
\ee
Inserting this equation into Eq.\ (\ref{SR.results.1}), we finally obtain
\be
\tilde\mathcal{S}(t,t_0) = \tilde\mathcal{R}(t,t_0)\,
  \left[1 + \int_{t_0}^t dt'\,
  \tilde\mathcal{R}(t',t_0)\right]^{-1} .
\label{SR.SRfinal}
\ee
This remarkable expression allows us to obtain the generator of the TCL Pauli
ME from the \textit{T}-matrix generator, in principle. This result is
potentially useful since we have an explicit expression for the transition
rates in the \textit{T}-matrix approach in terms of ordinary operators. It will
also allow us to derive the perturbative expansion of $\tilde\mathcal{S}(t,t_0)$
in the following.

The derivation also goes through for the full non-diagonal ME. The corresponding
expressions can be obtained by removing the tilde from all symbols. The result is
equivalent to an identity found by Bu\v{z}ek.\cite{Buz98}

\subsection{Perturbative expansion in the hybridization}
\label{sub.perb}

In this subsection we derive expansions of the TCL Pauli and
\textit{T}-matrix generators in powers of $H_\mathrm{hyb}$ or
$\Lh$. In the following, we send $t_0\to -\infty$ and suppress the arguments
$(t,-\infty)$. The expansion of the \textit{T}-matrix generator is obtained from
Eq.\ (\ref{A4.Rfull.4}), $\tilde\mathcal{R} = \sum_{\mu=1}^\infty
\tilde\mathcal{R}^{(2\mu)}$ with
\ba
\lefteqn{ \tilde\mathcal{R}^{(2\mu)} = -i\, e^{2\mu\eta t}\,
  \tP \Lh\, [-\LN + (2\mu-1)i\eta]^{-1}\, \Lh } \nonumber \\
&& {}\times [-\LN + (2\mu-2)i\eta]^{-1}\, \Lh \cdots 
  \Lh\,(-\LN+i\eta)^{-1} \nonumber \\
&& {}\times \Lh\tP . \hspace{10em}
\ea
We have used that all odd orders vanish.

The TCL generator is obtained from $\tilde\mathcal{R}$ using
Eq.\ (\ref{SR.SRfinal}). The time integral is easily performed,
\be
\tilde\mathcal{S} \equiv \sum_{\mu=1}^\infty
  \tilde\mathcal{S}^{(2\mu)}
  = \sum_{\mu=1}^\infty
  \tilde\mathcal{R}^{(2\mu)} \, \left[1 + \sum_{\mu=1}^\infty
  \frac{\tilde\mathcal{R}^{(2\mu)}}{2\mu\eta}\right]^{-1} .
\label{SR.Rexp}
\ee
Expanding the inverse and comparing the two sides order by order, we obtain
\be
\tilde\mathcal{S}^{(2\mu)} = \sum_{q=0}^{\mu-1}
  (-1)^q \!\! \sum_{\mu_0+\mu_1+\ldots+\mu_q = \mu} \!\!
  \tilde\mathcal{R}^{(2\mu_0)}\,
  \frac{\tilde\mathcal{R}^{(2\mu_1)}}{2\mu_1\eta} \cdots
  \frac{\tilde\mathcal{R}^{(2\mu_q)}}{2\mu_q\eta} ,
\label{A17.tS.3}
\ee
where the second sum is over $q+1$ positive integers $\mu_i$ adding
up to $\mu$. We note in passing that Eq.\ (\ref{A17.tS.3}) can also be obtained
from the expansion of the TCL generator in terms of ordered cumulants, following
van Kampen.\cite{Kam74,BrP02} The first few terms read explicitly
\ba
\tilde\mathcal{S}^{(2)} & = & \tilde\mathcal{R}^{(2)} , \\
\tilde\mathcal{S}^{(4)} & = & \tilde\mathcal{R}^{(4)}
  - \tilde\mathcal{R}^{(2)}\, \frac{\tilde\mathcal{R}^{(2)}}{2\eta} ,
\label{A17.tS4.3} \\
\tilde\mathcal{S}^{(6)} & = & \tilde\mathcal{R}^{(6)}
  - \tilde\mathcal{R}^{(4)}\, \frac{\tilde\mathcal{R}^{(2)}}{2\eta}
  - \tilde\mathcal{R}^{(2)}\, \frac{\tilde\mathcal{R}^{(4)}}{4\eta}
  \nonumber \\
&& {}+ \tilde\mathcal{R}^{(2)}\, \frac{\tilde\mathcal{R}^{(2)}}{2\eta}\,
    \frac{\tilde\mathcal{R}^{(2)}}{2\eta} .
\ea
The first equation shows that in
the sequential-tunneling approximation the TCL and \textit{T}-matrix
expressions for the transition rates agree.\cite{Tim08}

The problem in exploiting the expansion (\ref{A17.tS.3})
is that the $\tilde\mathcal{R}^{(2\mu)}$
diverge for $\eta\to 0^+$ for all $2\mu\ge 4$. This is in
addition to the explicit divergences due to negative powers of
$\eta$ in Eq.\ (\ref{A17.tS.3}). We would much prefer a representation
of $\tilde\mathcal{S}^{(2\mu)}$ in terms of expressions that remain finite.
To obtain one, we first simplify the notation by setting $t=0$,
since in the limit $\eta\to 0^+$ the value of $t$ does not matter.
We then define
\ba
\lefteqn{ \tilde\mathcal{R}^{(2\mu,2\mu')} := -i\, \tP\,
  \Lh\,
  [-\LN+(2\mu-1)i\eta]^{-1}\, \Lh } \nonumber \\
&& {}\times [-\LN+(2\mu-2)i\eta]^{-1}\, \Lh
  \cdots \Lh \nonumber \\
&& {}\times (-\LN+(2\mu'+1)i\eta)^{-1}\, \Lh\,
  \tP ,\hspace{5.5em}
\ea
where $\mu>\mu'$. Note the identity $\tilde\mathcal{R}^{(2\mu,0)} =
\tilde\mathcal{R}^{(2\mu)}$.

Divergences of the type of negative powers of $\eta$ arise
whenever $\LN$ in the inverse superoperators $(-\LN+i\kappa\eta)^{-1}$ can be
replaced by zero. These divergences are singled out by inserting
$1=\tP+\tQ$ between each pair of $\Lh$. We note that under the
assumption of non-degenerate dot states, the projection $\tP$ projects out
the \emph{secular reducible contributions}.\cite{KGL10} These are thus
removed by $\tQ$.
Since the lead-electron
creation and annihilation operators must be paired between any two $\tP$, all
expressions with an odd number of $\Lh$ superoperators between two $\tP$
projections vanish. Thus at the odd-numbered positions between the $\Lh$,
$\tQ=1-\tP$ does not do anything and $\tQ$ is redundant. This also
means that divergences cannot arise from the inverse superoperators
at odd-numbered positions. We therefore only insert $1=\tP+\tQ$ at the
even-numbered positions,
\ba
\lefteqn{ \tilde\mathcal{R}^{(2\mu)} = -i\, \tP\,
  \Lh\,
  [-\LN+(2\mu-1)i\eta]^{-1}\, \Lh\, (\tP+\tQ) } \nonumber \\
&& {}\times [-\LN+(2\mu-2)i\eta]^{-1}\, \Lh
  \cdots (\tP+\tQ)\, \Lh \nonumber \\
&& {}\times (-\LN+i\eta)^{-1}\,
  \Lh\, \tP .\hspace{12em}
\label{A13.Rmu.PQ.2}
\ea
We denote the regular parts of $\tilde\mathcal{R}^{(2\mu,2\mu')}$ by
\ba
\lefteqn{ \tilde\mathcal{R}^{(2\mu,2\mu')}_\mathrm{reg} := -i\, \tP\,
  \Lh\,
  [-\LN+(2\mu-1)i\eta]^{-1}\, \Lh\, \tQ } \nonumber \\
&& {}\times [-\LN+(2\mu-2)i\eta]^{-1}\, \Lh
  \cdots \tQ\, \Lh \nonumber \\
&& {}\times [-\LN+(2\mu'+1)i\eta]^{-1}\,
  \Lh\, \tP ,\hspace{7em}
\label{A13.Rregdef.3}
\ea
where a projection $\tQ$ is inserted at every even-numbered position between
the $\Lh$. We also define $\tilde\mathcal{R}^{(2\mu)}_\mathrm{reg}
:= \tilde\mathcal{R}^{(2\mu,0)}_\mathrm{reg}$. The finiteness of
$\tilde\mathcal{R}^{(2\mu,2\mu')}_\mathrm{reg}$
for $\eta\to 0^+$ is shown in a more general context
in appendix \ref{app.finite}. Note that
$\tilde\mathcal{R}^{(2\mu,2\mu-2)}_\mathrm{reg} =
\tilde\mathcal{R}^{(2\mu,2\mu-2)}$ and, in particular,
$\tilde\mathcal{R}^{(2)}_\mathrm{reg} = \tilde\mathcal{R}^{(2)}$, since there is
no position to insert $\tQ$. This reproduces the well-known observation that
the second-order rates in the \textit{T}-matrix formalism do not show
divergences.

From Eq.\ (\ref{A13.Rmu.PQ.2}) we now obtain, using Eq.\ (\ref{A4.Id.1a}),
\ba
\lefteqn{ \tilde\mathcal{R}^{(2\mu)} =
  \tilde\mathcal{R}^{(2\mu,0)}_\mathrm{reg}
  + \tilde\mathcal{R}^{(2\mu,2)}_\mathrm{reg}\,
    \frac{\tilde\mathcal{R}^{(2,0)}_\mathrm{reg}}{2\eta}
  + \tilde\mathcal{R}^{(2\mu,4)}_\mathrm{reg}\,
    \frac{\tilde\mathcal{R}^{(4,0)}_\mathrm{reg}}{4\eta} }
  \nonumber \\
&& {}+ \tilde\mathcal{R}^{(2\mu,4)}_\mathrm{reg}\,
    \frac{\tilde\mathcal{R}^{(4,2)}_\mathrm{reg}}{4\eta}\,
    \frac{\tilde\mathcal{R}^{(2,0)}_\mathrm{reg}}{2\eta}
  + \ldots \nonumber \\
&& {}+ \tilde\mathcal{R}^{(2\mu,2\mu-2)}_\mathrm{reg}\,
    \frac{\tilde\mathcal{R}^{(2\mu-2,2\mu-4)}_\mathrm{reg}}{(2\mu-2)\eta}\,
    \frac{\tilde\mathcal{R}^{(2\mu-4,2\mu-6)}_\mathrm{reg}}{(2\mu-4)\eta}
    \cdots \frac{\tilde\mathcal{R}^{(2,0)}_\mathrm{reg}}{2\eta} .
  \nonumber \\[-1ex]
&&
\label{A13.RRreg.3}
\ea
Since we have inserted $\tP+\tQ$ in $\mu-1$ positions, there are $2^{\mu-1}$
terms in this sum. In particular, we find
\ba
\tilde\mathcal{R}^{(2)} & = & \tilde\mathcal{R}^{(2,0)}_\mathrm{reg} , \\
\tilde\mathcal{R}^{(4)} & = & \tilde\mathcal{R}^{(4,0)}_\mathrm{reg}
  + \tilde\mathcal{R}^{(4,2)}_\mathrm{reg}\,
    \frac{\tilde\mathcal{R}^{(2,0)}_\mathrm{reg}}{2\eta} , \\
\tilde\mathcal{R}^{(6)} & = & \tilde\mathcal{R}^{(6,0)}_\mathrm{reg}
  + \tilde\mathcal{R}^{(6,2)}_\mathrm{reg}\,
    \frac{\tilde\mathcal{R}^{(2,0)}_\mathrm{reg}}{2\eta}
  + \tilde\mathcal{R}^{(6,4)}_\mathrm{reg}\,
    \frac{\tilde\mathcal{R}^{(4,0)}_\mathrm{reg}}{4\eta} \nonumber \\
&& {}+ \tilde\mathcal{R}^{(6,4)}_\mathrm{reg}\,
    \frac{\tilde\mathcal{R}^{(4,2)}_\mathrm{reg}}{4\eta}\,
    \frac{\tilde\mathcal{R}^{(2,0)}_\mathrm{reg}}{2\eta} .
\ea
As an intermediate result, we have thus written
the \textit{T}-matrix generator $\tilde\mathcal{R}$ order by order in terms of
expressions that remain finite for $\eta\to 0^+$ and explicit negative powers of
$\eta$. Since each insertion of $\tP$ generates a factor of
$1/\eta$, the most strongly diverging term in $\tilde\mathcal{R}^{(2\mu)}$
scales as $1/\eta^{\mu-1}$.

Inserting Eq.\ (\ref{A13.RRreg.3}) into
Eq.\ (\ref{A17.tS.3}), we obtain $\tilde\mathcal{S}^{(2\mu)}$ in
terms of $\tilde\mathcal{R}^{(2\nu,2\nu')}_\mathrm{reg}$ with $0\le\nu'<\nu\le
\mu$ for all $\mu$. The leading terms read
\ba
\tilde\mathcal{S}^{(2)} & = & \tilde\mathcal{R}^{(2,0)}_\mathrm{reg} , \\
\tilde\mathcal{S}^{(4)} & = & \tilde\mathcal{R}^{(4,0)}_\mathrm{reg}
  + \tilde\mathcal{R}^{(4,2)}_\mathrm{reg}\,
  \frac{\tilde\mathcal{R}^{(2,0)}_\mathrm{reg}}{2\eta}
  - \tilde\mathcal{R}^{(2,0)}_\mathrm{reg}\,
  \frac{\tilde\mathcal{R}^{(2,0)}_\mathrm{reg}}{2\eta}  ,
\label{A13.tS4.6} \\
\tilde\mathcal{S}^{(6)} & = & \tilde\mathcal{R}^{(6,0)}_\mathrm{reg}
  + \tilde\mathcal{R}^{(6,2)}_\mathrm{reg}\,
    \frac{\tilde\mathcal{R}^{(2,0)}_\mathrm{reg}}{2\eta}
  + \tilde\mathcal{R}^{(6,4)}_\mathrm{reg} \,
    \frac{\tilde\mathcal{R}^{(4,0)}_\mathrm{reg}}{4\eta} \nonumber \\
&& {}+ \tilde\mathcal{R}^{(6,4)}_\mathrm{reg} \,
    \frac{\tilde\mathcal{R}^{(4,2)}_\mathrm{reg}}{4\eta}\,
    \frac{\tilde\mathcal{R}^{(2,0)}_\mathrm{reg}}{2\eta}
  - \tilde\mathcal{R}^{(4,0)}_\mathrm{reg} \,
    \frac{\tilde\mathcal{R}^{(2,0)}_\mathrm{reg}}{2\eta} \nonumber \\
&& {}- \tilde\mathcal{R}^{(4,2)}_\mathrm{reg}\,
    \frac{\tilde\mathcal{R}^{(2,0)}_\mathrm{reg}}{2\eta}
    \, \frac{\tilde\mathcal{R}^{(2,0)}_\mathrm{reg}}{2\eta}
  - \tilde\mathcal{R}^{(2,0)}_\mathrm{reg}\,
    \frac{\tilde\mathcal{R}^{(4,0)}_\mathrm{reg}}{4\eta} \nonumber \\
&& {}- \tilde\mathcal{R}^{(2,0)}_\mathrm{reg}\,
  \frac{\tilde\mathcal{R}^{(4,2)}_\mathrm{reg}}{4\eta}\,
  \frac{\tilde\mathcal{R}^{(2,0)}_\mathrm{reg}}{2\eta}
  + \tilde\mathcal{R}^{(2,0)}_\mathrm{reg}\,
  \frac{\tilde\mathcal{R}^{(2,0)}_\mathrm{reg}}{2\eta}\,
    \frac{\tilde\mathcal{R}^{(2,0)}_\mathrm{reg}}{2\eta} .\nonumber \\[-1ex]
&&
\ea
In this expansion of the exact TCL Pauli generator, all singular
contributions in the limit $\eta\to 0^+$ have been made explicit. The maximum
power is $1/\eta^{\mu-1}$.

To conclude this section, we illustrate the results by considering the terms
of fourth order. The corresponding term in the \textit{T}-matrix generator
reads
\ba
\lefteqn{ \tilde\mathcal{R}^{(4)} = -i\,
  \tP \Lh\, (-\LN + 3i\eta)^{-1}\, \Lh } \nonumber \\
&& {}\times (-\LN + 2i\eta)^{-1}\, \Lh \,(-\LN+i\eta)^{-1}\, \Lh\tP .
\label{A13.R4.1}
\ea
Let $\tilde\mathcal{R}^{(4)}$ act upon some density operator $\rho$. Then $\Lh
\,(-\LN+i\eta)^{-1}\, \Lh\tP\rho$ contains contributions for which the second
(from the right) superoperator $\Lh$ undoes the changes introduced by the first
$\Lh$. Hence, $\Lh \,(-\LN+i\eta)^{-1}\, \Lh\tP\rho$ is an operator with
non-vanishing \emph{diagonal} components in the product basis of unperturbed
eigenstates. But for diagonal components $|j\rangle\langle j|$ we have
$\LN\,|j\rangle\langle j|=0$ so that $\LN$ in the next superoperator to the
left, $(-\LN + 2i\eta)^{-1}$, can be replaced by zero. We thus obtain a singular
contribution proportional to $1/2i\eta$. More
formally, we single out the divergent contributions by introducing $1=\tQ+\tP$,
\ba
\lefteqn{ \tilde\mathcal{R}^{(4)}
  = -i\, \tP \Lh\, (-\LN + 3i\eta)^{-1}\, \Lh\,
  (-\LN + 2i\eta)^{-1} } \nonumber \\
&& \quad{}\times \tQ \Lh \,(-\LN+i\eta)^{-1}\, \Lh\tP \nonumber \\
&& {}-i\, \tP \Lh\, (-\LN + 3i\eta)^{-1}\, \Lh\,
  \frac{-i}{2\eta} \nonumber \\
&& \quad{}\times \tP \Lh \, (-\LN+i\eta)^{-1}\, \Lh\tP \nonumber \\
&& =: \: \tilde\mathcal{R}^{(4)}_\mathrm{reg}
  + \tilde\mathcal{R}^{(4)}_\mathrm{div} . \hspace{14em}
\label{A13.R4.2}
\ea
The divergent part $\tilde\mathcal{R}^{(4)}_\mathrm{div}$ is
identical to $\tilde\mathcal{R}^{(4,2)}_\mathrm{reg}\,
\tilde\mathcal{R}^{(2,0)}_\mathrm{reg}/2\eta$, according to the definition
(\ref{A13.Rregdef.3}).

The fourth-order term $\tilde\mathcal{S}^{(4)}$ of the TCL generator contains a
correction term beyond $\tilde\mathcal{R}^{(4)}$, cf.\ Eq.\ (\ref{A17.tS4.3}),
namely
\ba
\lefteqn{ - \tilde\mathcal{R}^{(2)}\, \frac{\tilde\mathcal{R}^{(2)}}{2\eta}
  \equiv - \tilde\mathcal{R}^{(2)}_\mathrm{reg}\,
  \frac{\tilde\mathcal{R}^{(2)}_\mathrm{reg}}{2\eta} } \nonumber \\
&& = +i\, \tP \Lh \,
  (-\LN + i\eta)^{-1}\, \Lh \,
  \frac{-i}{2\eta} \nonumber \\
&& \quad{}\times \tP \Lh \,(-\LN+i\eta)^{-1}\, \Lh\tP .\hspace{5em}
\ea
This looks very similar to the divergent part
$\tilde\mathcal{R}^{(4)}_\mathrm{div}$. The differences
are the opposite sign and a different prefactor of $i\eta$ in
the left-most inverse superoperator. If this factor were the same, the
correction term would exactly cancel the divergent part. As it is, the
correction term does remove the divergence for $\eta\to 0^+$ but leaves a
non-zero difference behind,
\ba
\lefteqn{ \tilde\mathcal{R}^{(4)}_\mathrm{div} - \tilde\mathcal{R}^{(2)}\,
  \frac{\tilde\mathcal{R}^{(2)}}{2\eta} = i\, \tP \Lh \,
  (-\LN + i\eta)^{-1}
  (-\LN + 3i\eta)^{-1} } \nonumber \\
&& {}\times \Lh \tP \Lh \, (-\LN+i\eta)^{-1}
  \Lh \tP .\hspace{8em}
\label{A4.divcan.2}
\ea
We will show that this difference indeed remains finite.

\subsection{Cancelation of divergences}
\label{sub.cancel}

Our next goal is to show that the divergences described by negative powers of $\eta$
all cancel in the limit $\eta\to 0^+$. It is useful to resum the terms in Eq.\
(\ref{A17.tS.3}),
\be
\tilde\mathcal{S} = \sum_{q=0}^\infty (-1)^q
  \! \sum_{\mu_0,\mu_1,\ldots,\mu_q=1}^\infty \!
  \tilde\mathcal{R}^{(2\mu_0)}\,
  \frac{\tilde\mathcal{R}^{(2\mu_1)}}{2\mu_1\eta} \cdots
  \frac{\tilde\mathcal{R}^{(2\mu_q)}}{2\mu_q\eta} .
\ee
Inserting Eq.\ (\ref{A13.RRreg.3}), we obtain
\ba
\tilde\mathcal{S} & = & \sum_{p=0}^\infty
  \sum_{\mu_0,\mu_0',\mu_1,\mu_1',\mu_2,\mu_2',\ldots,\mu_p} \!\!
  (-1)^{n'}\, \tilde\mathcal{R}^{(2\mu_0,2\mu_0')}_\mathrm{reg}\,
  \frac{\tilde\mathcal{R}^{(2\mu_1,2\mu_1')}_\mathrm{reg}}{2\mu_1\eta}
  \nonumber \\
&& {}\times \frac{\tilde\mathcal{R}^{(2\mu_2,2\mu_2')}_\mathrm{reg}}{2\mu_2\eta}
  \cdots
  \frac{\tilde\mathcal{R}^{(2\mu_p,2\mu_p'=0)}_\mathrm{reg}}{2\mu_p\eta} ,
\label{A17.Sexpreg.4}
\ea
where $n'+1$ is the number of $\mu_i'$ being zero. The second sum is over $p+1$
pairs $(\mu_i,\mu_i')$, $i=0,1,\ldots,p$, with $\mu_i=1,2,\ldots$,
$\mu_i'=0,1,\ldots$, and $\mu_i>\mu_i'$, satisfying either $\mu_i' = \mu_{i+1}$
or $\mu_i' = 0$ for any two consecutive pairs. The last $\mu_i'=\mu_p'$ must
equal zero.

In Eq.\ (\ref{A17.Sexpreg.4}), $p$ represents the explicit order
in $1/\eta$. However, the superoperators
$\tilde\mathcal{R}^{(2\mu,2\mu')}_\mathrm{reg}$ also depend on $\eta$.
To find the limit $\eta\to 0^+$, we thus have to expand them
up to the order $\eta^p$. Their Taylor series in $\eta$ reads
\ba
\lefteqn{ \tilde\mathcal{R}^{(2\mu,2\mu')}_\mathrm{reg} = -i
  \sum_{m_{2\mu-1},m_{2\mu-2},\ldots,m_{2\mu'+1}=0}^\infty } \nonumber \\
&& {}\times (-i\,\eta)^{m_{2\mu-1}+m_{2\mu-2}+\ldots+m_{2\mu'+1}} \nonumber \\
&& {}\times (2\mu-1)^{m_{2\mu-1}} (2\mu-2)^{m_{2\mu-2}} \cdots
  (2\mu'+1)^{m_{2\mu'+1}} \nonumber \\
&& {}\times [m_{2\mu-1},m_{2\mu-2},\ldots,m_{2\mu'+1}]^{(2\mu,2\mu')} ,
\label{A17.Rregex.5}
\ea
where we have defined the notation
\ba
\lefteqn{ [m_{2\mu-1},m_{2\mu-2},\ldots,m_{2\mu'+1}]^{(2\mu,2\mu')}
  := \lim_{\eta\to 0^+} \tP\,\Lh } \nonumber \\
&& {}\times   [-\LN+(2\mu-1)i\eta]^{-1-m_{2\mu-1}}\,
  \Lh\,\tQ \nonumber \\
&& {}\times [-\LN+(2\mu-2)i\eta]^{-1-m_{2\mu-2}}\,
  \Lh \cdots \Lh \nonumber \\
&& {}\times [-\LN+(2\mu'+1)i\eta]^{-1-m_{2\mu'+1}}\, \Lh\, \tP \hspace{4em}
\label{A17.brackdef.3}
\ea
with $\tQ$ inserted at all even-numbered positions. In particular, Eq.\
(\ref{A17.Rregex.5}) implies that
\be
\lim_{\eta\to 0^+} \tilde\mathcal{R}^{(2\mu)}_\mathrm{reg}
  = -i\, [0,0,\ldots,0]^{(2\mu,0)} .
\label{A17.Rregbrack.2}
\ee
It is shown in appendix \ref{app.finite} that the limit $\eta\to 0^+$ in Eq.\
(\ref{A17.brackdef.3}) converges for all
$m_{2\mu-1},m_{2\mu-2},\ldots,m_{2\mu'+1}\ge 0$.
Moreover, we show in appendix \ref{app.indep} that the superoperator defined in Eq.\ (\ref{A17.brackdef.3}) does not
depend on the values of the prefactors of $i\eta$, as long as they are all
positive. Thus it does
not depend on $\mu$ and $\mu'$ except in so far as $2\mu-2\mu'-1$ is the number
of its arguments $m_\nu$. We therefore drop the superscript $(2\mu,2\mu')$ from
now on.

Insertion of Eq.\ (\ref{A17.Rregex.5}) into Eq.\ (\ref{A17.Sexpreg.4}) leads to
an expansion of the TCL generator $\tilde\mathcal{S}$,
\ba
\lefteqn{ \tilde\mathcal{S} =
  \sum_{p=0}^\infty\, (-i)^{p+1} \!\!
  \sum_{\mu_0,\mu_0',\mu_1,\mu_1',\mu_2,\mu_2',\ldots\,\mu_p} \!\!
  (-1)^{n'} } \nonumber \\
&& {}\times \sum_{m_{0,2\mu_0-1},\ldots,m_{0,2\mu_0'+1},
  \ldots,m_{p,2\mu_p-1},\ldots,m_{p,1}=0}^\infty \nonumber \\
&& {}\times \frac{(-i)^{\Sigma_m}}{2\mu_1\,2\mu_2 \cdots 2\mu_p}\,
  (2\mu_0-1)^{m_{0,2\mu_0-1}} \cdots \nonumber \\
&& {}\times (2\mu'_0+1)^{m_{0,2\mu_0'+1}} \cdots
  (2\mu_p-1)^{m_{p,2\mu_p-1}} \cdots \nonumber \\
&& {}\times [m_{0,2\mu_0-1},\ldots,m_{0,2\mu'_0+1}] \cdots
  [m_{p,2\mu_p-1},\ldots,m_{p,1}] \nonumber \\
&& {}\times \eta^{\Sigma_m-p} , \hspace{8em}
\label{A17.Sexpbb.3}
\ea
where $\Sigma_m := m_{0,2\mu_0-1}+\ldots+m_{p,1}$ is the sum of all $m_{i,\nu}$.
The two indices of $m_{i,\nu}$ enumerate the factors of
$\tilde\mathcal{R}^{(2\mu,2\mu')}_\mathrm{reg}$ in Eq.\ (\ref{A17.Sexpreg.4})
and the inverse superoperators in
$[m_{2\mu-1},m_{2\mu-2},\ldots,m_{2\mu'+1}]$, respectively.

Terms containing positive powers $\Sigma_m-p>0$ of $\eta$ vanish in the limit
$\eta\to 0^+$ and can thus be disregarded. On the other hand, to obtain a finite
limit, the prefactors in all terms involving negative powers $\Sigma_m-p<0$ must
cancel. The cancelations can only involve terms with the same superoperator
factor $[m_{0,2\mu_0-1},\ldots,m_{0,2\mu'_0+1}] \cdots
[m_{p,2\mu_p-1},\ldots,m_{p,1}]$. These terms have the same values of $p$, of
the orders $2n_i=2\mu_i-2\mu_i'$, and of all $m_{i,\nu}$. We thus write
\ba
\lefteqn{\tilde\mathcal{S} =
  \sum_{p=0}^\infty\, (-i)^{p+1}
  \sum_{n_0,n_1,\ldots,n_p=1}^\infty } \nonumber \\
&& {}\times \sum_{m_{0,2\mu_0-1},\ldots,m_{0,2\mu_0'+1},
  \ldots,m_{p,2\mu_p-1},\ldots,m_{p,1}=0}^\infty 
  (-i)^{\Sigma_m} \nonumber \\
&& {}\times f(n_0,n_1,\ldots,n_p;m_{0,2\mu_0-1},\ldots,m_{p,1})
  \nonumber \\
&& {}\times [m_{0,2\mu_0-1},\ldots,m_{0,2\mu'_0+1}] \cdots
  [m_{p,2\mu_p-1},\ldots,m_{p,1}] \nonumber \\
&& {}\times \eta^{\Sigma_m-p} \hspace{9em}
\label{A17.Sexpbb.4}
\ea
with the prefactors
\ba
\lefteqn{ f(n_0,n_1,\ldots,n_p;m_{0,2\mu_0-1},\ldots,m_{p,1}) }
  \nonumber \\
&& := \!\!\! \sum_{\mu_0,\mu_0',\mu_1,\mu_1',\mu_2,\mu_2',\ldots,\mu_p}
  \!\! \frac{(-1)^{n'} }{2\mu_1\,2\mu_2 \cdots 2\mu_p} \,
  (2\mu_0-1)^{m_{0,2\mu_0-1}} \nonumber \\
&& {}\times \cdots (2\mu'_0+1)^{m_{0,2\mu_0'+1}} \cdots
  (2\mu_p-1)^{m_{p,2\mu_p-1}} \cdots  ,\hspace{0.5em}
\label{A17.sumint.4}
\ea
where the sum in Eq.\ (\ref{A17.sumint.4})
is now constrained by $2\mu_i-2\mu_i'=2n_i$ being given.
With this constraint, the only freedom left in the sum is the choice of which
$\mu_i'$ are zero; recall that the non-zero
$\mu_i'$ equal $\mu_{i+1}$. The numbers
$\mu_0,\mu_0',\mu_1,\mu_1',\mu_2,\mu_2',\ldots,\mu_p$ can be reconstructed from
the orders $n_0,n_1,n_2,\ldots,n_p$ and the indices $i$ of the $\mu_i'$ that
equal zero. Defining the set
\be
Z := \{i|\mu_i'=0\} ,
\ee
we have $p\in Z$ and $n'=|Z|-1$, where $|Z|$ is the cardinality of $Z$.
Defining the ``non-member function''
\be
\pi^Z_i := \left\{\begin{array}{ll}
  0 & \mbox{if $i\in Z$} \\
  1 & \mbox{if $i\notin Z$}
  \end{array}\right.
\ee
we have $\pi^Z_p=0$ and $n' = p-\sum_{i=0}^{p-1} \pi^Z_i$.
Replacing $\mu_i'$ by $\mu_i-n_i$ in Eq.\ (\ref{A17.sumint.4}) we obtain
\ba
\lefteqn{
f(n_0,n_1,\ldots,n_p;m_{0,2\mu_0-1},\ldots,m_{p,1}) } \nonumber \\
&& = (-1)^p \!\!
  \sum_{\pi^Z_0,\pi^Z_1,\ldots,\pi^Z_{p-1}=0}^1
  \frac{\prod_{i=0}^{p-1} \, (-1)^{\pi^Z_i}}
  {2\mu_1 2\mu_2 \cdots 2\mu_p} \nonumber \\
&& {}\times (2\mu_0-1)^{m_{0,2\mu_0-1}} (2\mu_0-2)^{m_{0,2\mu_0-2}}\cdots
  \nonumber \\
&& {}\times (2\mu_0-2n_0+1)^{m_{0,2\mu_0-2n_0+1}} \cdots \nonumber \\
&& {}\times (2\mu_p-1)^{m_{p,2\mu_p-1}} (2\mu_p-2)^{m_{p,2\mu_p-2}}\cdots
  1^{m_{p,1}} , \hspace{1em}
\label{A17.f.3}
\ea
where
$\mu_i = n_i + \pi^Z_i n_{i+1} + \pi^Z_i \pi^Z_{i+1} n_{i+2} + \ldots$
and $\mu_p = n_p$, and products are understood to equal unity if they do not
contain any factors. For convenience, we define
\be
M_i := m_{i,2\mu_i-1} + m_{i,2\mu_i-2} + \ldots + m_{i,2\mu_i-2n_i+1} .
\ee
Note that $\Sigma_m=\sum_{i=0}^p M_i$.

The evaluation of Eq.\ (\ref{A17.f.3}), which is presented in appendix
\ref{app.f}, is rather lengthy but has a remarkably
simple result: For all $\Sigma_m\le p$,
\be
f(n_0,n_1,\ldots,n_p;m_{0,2\mu_0-1},\ldots,m_{p,1}) = 1
\ee
if and only if
\be
\Sigma_m-p=M_0+M_1+\ldots+M_p-p=0
\ee
and there does not exist any integer $i<p$ such that
\be
M_0+M_1+\ldots+M_i-i = 0 .
\ee
Otherwise, $f=0$. Recall that $f$ is not of interest if $\Sigma_m>p$.
Furthermore, we also show in appendix \ref{app.f} that the condition for
non-zero $f$ can only be satisfied if $m_{p,2\mu_p-1}=\ldots=m_{p,1}=0$.

Inserting these results into Eq.\ (\ref{A17.Sexpbb.4}), taking the limit
$\eta\to 0^+$, and renumbering the $m_{i,\nu}$, we obtain
\ba
\lefteqn{ \lim_{\eta\to 0^+} \tilde\mathcal{S} =
  -i \sum_{p=0}^\infty\, (-1)^p
  \sum_{n_0,n_1,\ldots,n_p=1}^\infty } \nonumber \\
&& {}\times \sum_{m_{0,1},\ldots,m_{0,2n_0-1},
  \ldots,m_{p-1,1},\ldots,m_{p-1,2n_{p-1}-1}} 
  \nonumber \\
&& {}\times \Theta(M_0-1)\, \Theta(M_0+M_1-2) \cdots \nonumber \\
&& {}\times \Theta(M_0+M_1+\ldots+M_{p-1}-p) 
  \nonumber \\
&& {}\times [m_{0,1},\ldots,m_{0,2n_0-1}] \cdots
  [m_{p-1,1},\ldots,m_{p-1,2n_{p-1}-1}] \nonumber \\
&& {}\times [0,\ldots,0] ,
\label{A17.Sexp.20}
\ea
where the sum over the $m_{i,\nu}$ is constrained by
$m_{0,1}+\ldots+m_{p-1,2n_{p-1}-1}=p$ and we have defined
\be
\Theta(n) := \left\{\begin{array}{ll}
  0 & \mbox{for $n<0$,} \\
  1 & \mbox{for $n\ge 0$.}
  \end{array}\right.
\ee
Note that the factor $\Theta(M_0+M_1+\ldots+M_{p-1}-p)$ is redundant.

The $p=0$ contribution in Eq.\ (\ref{A17.Sexp.20}) does not contain any sums over
$m_{i,\nu}$ since $\Sigma_m=p=0$. There just remains a sum over $n_0$, the order
in $H_\mathrm{hyb}$, i.e., the $p=0$ contribution reads $-i\,[0] -i\, [0,0,0] -
\ldots$ According to Eq.\ (\ref{A17.Rregbrack.2}), this equals $\lim_{\eta\to
0^+} \sum_\mu \tilde\mathcal{R}^{(2\mu)}_\mathrm{reg}$. Thus in the expansion in
$H_\mathrm{hyb}$, the expansion term $\tilde\mathcal{S}^{(2\mu)}$ of the TCL
generator contains the properly regularized \textit{T}-matrix term
$\tilde\mathcal{R}^{(2\mu)}_\mathrm{reg}$ plus corrections. Furthermore, all
these corrections contain $[0,\ldots,0]$, i.e., an expansion term of
$\tilde\mathcal{R}_\mathrm{reg}$, as the right-most superoperator factor.

Suppressing the limit directive from now on,
we find that $i\,\tilde\mathcal{S}$ in Eq.\ (\ref{A17.Sexp.20}) is the
sum of all terms that can be constructed according to the following rules:
\begin{enumerate}
\item Each term is a product of $p+1=1,2,\ldots$ superoperators of the form
$[m_{j,1},\ldots,m_{j,2n_j-1}]$ with $j=0,\ldots,p$, $n_j=1,2,\ldots$, and
$m_{j,\nu}=0,1,\ldots$
\item Defining $M_j := m_{j,1}+\ldots m_{j,2n_j-1}$, only terms with
$M_0+M_1+\ldots+M_j>j$ for all $j<p$ are allowed.
\item Only terms with $M_0+M_1+\ldots+M_p=p$ are allowed.
\item Each term obtains a factor $(-1)^p$.
\end{enumerate}
We draw a number of conclusions: If an allowed term contains a factor
$[m_{j,1},\ldots,m_{j,2n_j-1}]$ then any term with this factor replaced by
$[m'_{j,1},\ldots,m'_{j,2n_j'-1}]$ with $m'_{j,1}+\ldots+m'_{j,2n_j'-1} =
m_{j,1}+\ldots+m_{j,2n_j-1}$ is also allowed.
If we denote the sum of all such terms by
\be
[[M]] := \sum_{n=1}^\infty \, \sum_{m_1+\ldots+m_{2n-1}=M}
  [m_1,\ldots,m_{2n-1}] ,
\label{A17.bbMdef.1}
\ee
we obtain
\ba
\lefteqn{ i\,\tilde\mathcal{S} = \sum_{p=0}^\infty \, (-1)^p
  \sum_{M_0=1}^\infty \, \sum_{M_1=\max(0,2-M_0)}^\infty
  } \nonumber \\
&& {}\times \sum_{M_2=\max(0,3-M_0-M_1)}^\infty \cdots
  \sum_{M_{p-1}=\max(0,p-M_0-\ldots-M_{p-2})}^{p-M_0-M_1-\ldots-M_{p-2}}
  \nonumber \\
&& {}\times [[M_0]]\, [[M_1]] \cdots [[M_{p-1}]]\, [[0]] .\hspace{9em}
\label{A17.Sfull.10}
\ea
The last sum is understood to equal zero if the upper limit is smaller than the
lower one. In order to obtain an expansion of
$i\tilde\mathcal{S}$ in powers of $H_\mathrm{hyb}$, we note that
$[m_1,\ldots,m_{2n-1}]$ is of order
$H_\mathrm{hyb}^{2n}$ with $2n\ge 2$. Thus $[[M]]$ contains contributions of
second and higher orders. To obtain the expansion term
$i\tilde\mathcal{S}^{(2\mu)}$ of order $2\mu$, we thus only have to consider
terms with $p+1\le \mu$ factors $[[M_j]]$
in Eq.\ (\ref{A17.Sfull.10}). The first few terms read
\ba
\tilde\mathcal{S}^{(2)} & = & -i\, [0] ,
\label{A17.S2.exp} \\
\tilde\mathcal{S}^{(4)} & = & -i\, [0,0,0] + i\, [1][0] ,
\label{A17.S4.exp} \\
\tilde\mathcal{S}^{(6)} & = & -i\, [0,0,0,0,0] \nonumber \\
&& {}+ i\, [0,0,1][0] + i\, [0,1,0][0] + i\, [1,0,0][0] \nonumber \\
&& {}+ i\, [1][0,0,0] - i\, [1][1][0] - i\, [2][0][0] .
\label{A17.S6.exp}
\ea
Higher-order terms are easily generated using computer algebra.
They become increasingly lengthy; $\tilde\mathcal{S}^{(8)}$
contains $30$ terms and $\tilde\mathcal{S}^{(10)}$ already $143$.
Simplification is possible by realizing that some of the terms in
$\tilde\mathcal{S}^{(2\mu)}$ contain factors of $\tilde\mathcal{S}^{(2\mu')}$
with $\mu'<\mu$, as we show now.

Equation (\ref{A17.Sfull.10}) is equivalent to the surprising identity
\be
i\,\tilde\mathcal{S} = \sum_{M=0}^\infty \, [[M]] \,
  \big({-i}\,\tilde\mathcal{S}\big)^M .
\label{A17.Seqfinal.3}
\ee
The usefulness of this equation rests on the observation that $[[M]]$ is of at
least second order in $H_\mathrm{hyb}$. Therefore, we can express
$\tilde\mathcal{S}^{(2\mu)}$ by \emph{lower-order} terms
$\tilde\mathcal{S}^{(2\mu')}$, $\mu'<\mu$. Together with the starting value
$\tilde\mathcal{S}^{(2)} = -i\, [0] = \tilde\mathcal{R}^{(2)}_\mathrm{reg}$,
we obtain a recursive scheme for determining $\tilde\mathcal{S}^{(2\mu)}$.

To prove that Eq.\ (\ref{A17.Seqfinal.3}) has the solution given in Eq.\
(\ref{A17.Sfull.10}), we iterate Eq.\ (\ref{A17.Seqfinal.3}),
\ba
i\,\tilde\mathcal{S} & = & [[0]] + \sum_{M_0=1}^\infty [[M_0]]\,
  (-i\,\tilde\mathcal{S})^{M_0} \nonumber \\
& = & [[0]] - \sum_{M_0=1}^\infty \sum_{M_1=0}^\infty
  [[M_0]]\, [[M_1]]\, (-i\,\tilde\mathcal{S})^{M_0+M_1-1} \nonumber \\
& = & [[0]] - [[1]]\,[[0]]
  + \sum_{M_0=1}^\infty \sum_{M_1=\max(0,2-M_0)}^\infty
  \sum_{M_2=0}^\infty [[M_0]] \nonumber \\
&& {}\times [[M_1]]\, [[M_2]]\,
  (-i\tilde\mathcal{S})^{M_0+M_1+M_2-2} \nonumber \\
& = & [[0]] - [[1]]\,[[0]]
  + \sum_{M_0=1}^2 [[M_0]]\, [[2-M_0]]\, [[0]] \mp \ldots \nonumber \\[-1ex]
&&
\ea
It is clear how this continues. The terms no longer containing
$\tilde\mathcal{S}$ are the ones satisfying the conditions in the multiple sum
in Eq.\ (\ref{A17.Sfull.10}). Thus Eq.\ (\ref{A17.Sfull.10}) is a solution of
Eq.\ (\ref{A17.Seqfinal.3}). To show that it is the only solution, i.e., that
Eq.\ (\ref{A17.Sfull.10}) implies Eq.\ (\ref{A17.Seqfinal.3}), we note that
the iteration shows that any solution $\tilde\mathcal{S}'$ of Eq.\
(\ref{A17.Seqfinal.3}) agrees with Eq.\ (\ref{A17.Sfull.10}) order by order in
the number $p+1$ of superoperator factors. Thus we find
$\tilde\mathcal{S}'=\tilde\mathcal{S}$ to any order $p+1$.

\subsection{Relation to the Nakajima-Zwanzig master equation}
\label{sub.NZ}

The Nakajima-Zwanzig (NZ) ME\cite{Nak58,Zwa60} and equivalent formulations are
commonly used in the field of transport through nanostructures.
The real-time diagrammatic technique\cite{ScS94,KSS95,KSS96} and the suitably
generalized Wangsness-Bloch-Redfield theory\cite{WaB53,Blo57,Red65} are
such equivalent formulations.\cite{Tim08,KGL10} We again only
consider the initial condition $\mathcal{Q}\rho(t_0)=0$. The derivation starts
in the same way as the one of the TCL ME, leading to Eq.\
(\ref{A2.sol2.2}). Inserting this equation into Eq.\ (\ref{A2.vN.2a}) and
using the identities (\ref{A2.Id.1}) and (\ref{A2.Id.2}), we obtain
the NZ ME\cite{BrP02}
\ba
\lefteqn{ \frac{d}{dt}\, \mathcal{P}\rho(t) =
  -i \LN \mathcal{P}\rho(t) 
  - \mathcal{P} \Lh e^{\eta t}
  \int_{t_0}^t dt' } \nonumber \\
&& {}\times T_\leftarrow \exp\left[-i\int_{t'}^t dt'' \,
  (\LN + \mathcal{Q}\Lh e^{\eta t''})\right] \nonumber \\
&& {}\times \mathcal{Q} \Lh e^{\eta t'} \mathcal{P} \rho(t') .
\label{NZ.NZ.5}
\ea
Expansion in powers of $\Lh$ yields
\ba
\lefteqn{ \frac{d}{dt}\, \mathcal{P}\rho(t) =
  -i \LN \mathcal{P}\rho(t)
  - \mathcal{P} \Lh e^{\eta t}
  \sum_{\nu=0}^\infty (-i)^\nu } \nonumber \\
&& {}\times \int_{t_0}^t dt_1
  \int_{t_0}^{t_1} dt_2
  \cdots \int_{t_0}^{t_{\nu-1}} dt_\nu
  \int_{t_0}^{t_\nu} dt_{\nu+1}\,
  e^{-i\LN\,(t-t_1)} \nonumber \\
&& {}\times \mathcal{Q}\Lh e^{\eta t_1} \,
  e^{-i\LN\,(t_1-t_2)}
  \mathcal{Q}\Lh e^{\eta t_2} \cdots
  \mathcal{Q}\Lh e^{\eta t_\nu} \nonumber \\
&& {}\times e^{-i\LN\,(t_\nu-t_{\nu+1})}
  \mathcal{Q}
  \Lh e^{\eta t_{\nu+1}} \mathcal{P} \rho(t_{\nu+1}) .
\label{NZ.NZ.7}
\ea
As above, the projections $\mathcal{Q}$ at odd-numbered positions are redundant,
while at even-numbered positions they remove divergent reducible
contributions.\cite{Tim08,BKG10,KGL10}

The derivation goes through if we replace $\mathcal{P}$ and $\mathcal{Q}$ by
$\tP$ and $\tQ$, respectively. We end up with a
Nakajima-Zwanzig-Pauli ME for the diagonally projected density operator,
\ba
\lefteqn{ \frac{d}{dt}\, \tP\rho(t) = - \tP \Lh e^{\eta t}
  \int_{t_0}^t dt'\, T_\leftarrow \exp\bigg[-i\int_{t'}^t dt'' }
  \nonumber \\
&& {}\times (\LN + \tQ\Lh e^{\eta t''})\bigg] \tQ
  \Lh e^{\eta t'} \tP \rho(t') .\hspace{3em}
\label{NZ.NZ.9}
\ea
The bare time evolution has dropped out because of Eq.\
(\ref{A4.Id.1a}). It is this ME that is expanded up to fourth order
in Refs.\ \onlinecite{LeW08} and \onlinecite{KGL10}.
The projections $\tQ$ now remove only the
\emph{diagonal} reducible contributions, not all of them. They thus implement
the regularization discussed by Koller \textit{et al.}\cite{KGL10} As in
Ref.\ \onlinecite{KGL10}, the regularization is automatically included.
Our result shows that it can be formulated compactly using
suitable projection operators $\tP$ and $\tQ$. It has been noted in Ref.\
\onlinecite{BKG10} and shown explicitly in Ref.\ \onlinecite{KGL10}
that the Turek-Matveev scheme\cite{TuM02,KOO04} differs from
this built-in regularization already at fourth order.

If one is only interested in the stationary solution of the ME, $\tP \rho(t')$
on the right-hand side of Eq.\ (\ref{NZ.NZ.9}) can be taken to be
time-independent. It is then possible to evaluate the time integrals explicitly.
The resulting equation for the stationary state reads
$0 = \tilde\mathcal{G}\,\tP\rho$ with the generator, for $t_0\to -\infty$,
\ba
\lefteqn{ \tilde\mathcal{G} \equiv
  \sum_{\mu=1}^\infty \tilde\mathcal{G}^{(2\mu)}
  := - i \tP \Lh
  \sum_{\mu=1}^\infty
  e^{2\mu\eta t} } \nonumber \\
&& {}\times [-\LN+(2\mu-1)i\eta]^{-1} \,
  \Lh\tQ \,
  [-\LN+(2\mu-2)i\eta]^{-1} \nonumber \\
&& {}\times \Lh \cdots
  \tQ\Lh \, (-\LN+i\eta)^{-1}\, \Lh\, \tP . \hspace{5em}
\label{NZ.Gdef.3}
\ea
The redundant projections $\tQ$ at odd-numbered positions have been omitted.
Since this is an exact result for the stationary state, it should agree with
what the exact TCL ME predicts. We will return to this point shortly.

If one is interested in the dynamics, one can still obtain a local ME
from the NZ approach. This requires the Markov approximation, which is
based on the assumption that the memory
kernel in Eq.\ (\ref{NZ.NZ.9}) decays rapidly in time. This
assumption is often justified since relaxation in the leads is rapid but also
follows directly from the condition of a nearly closed conduction channel,
$I/V \ll e^2/h$.\cite{Tim08} With the Markov approximation,
$\tP\rho(t')$ is replaced by
$\tP\rho(t)$. Taking $t_0\to-\infty$, one obtains the approximate
``Nakajima-Zwanzig-Markov-Pauli'' ME
\be
d\tP\rho(t)/dt = \tilde\mathcal{G}\, \tP\rho(t)
\ee
with the generator $\tilde\mathcal{G}$ defined in Eq.\ (\ref{NZ.Gdef.3}).

Comparison of Eq.\ (\ref{NZ.Gdef.3}) and Eq.\ (\ref{A13.Rregdef.3}) shows that
the expansion terms are identical to the \emph{properly regularized} expansion
terms of the \textit{T}-matrix generator (we suppress the limit $\eta\to0^+$),
\be
\tilde\mathcal{G}^{(2\mu)} = \tilde\mathcal{R}^{(2\mu)}_\mathrm{reg} .
\label{NZ.GRreg.2}
\ee
Hence, the Nakajima-Zwanzig-Markov-Pauli
ME is identical, order by order in $H_\mathrm{hyb}$, to the ME
with rates obtained from the \textit{T}-matrix approach and regularized
by dropping secular reducible contributions. Up to fourth order, this has
been shown by Koller \textit{et al.}\cite{KGL10}

We can now gain additional insight into the failure\cite{KGL10} of
the Turek-Matveev regularization scheme.\cite{TuM02}
The proper regularization of the
\textit{T}-matrix expressions can be understood as omitting all terms in Eq.\
(\ref{A13.RRreg.3}) except for the first one or, in other words, as omitting all terms
containing \emph{explicit}
negative powers of $\eta$. The Turek-Matveev scheme, applied to
the calculation of the fourth-order rates,\cite{KOO04} corresponds to expanding the
rates into powers of $\eta$ and omitting the diverging part proportional to $1/\eta$
and then letting $\eta$ go to zero. The obvious generalization to all orders is to omit all
negative powers of $\eta$. The two regularization
procedures thus look quite similar. They are not identical, though, since the
superoperators $\tilde\mathcal{R}^{(2\mu,2\mu')}_\mathrm{reg}$ appearing in the
proper expansion (\ref{A13.RRreg.3}) contain positive powers of $\eta$. The
positive powers from $\tilde\mathcal{R}^{(2\mu,2\mu')}_\mathrm{reg}$ together with the
explicit negative powers lead to terms of order $\eta^0$, which are retained by the
Turek-Matveev scheme but are absent in the proper regularization. We reiterate
that both the Nakajima-Zwanzig-Markov-Pauli ME and the TCL Pauli ME are
automatically regularized---for the TCL case this is one of our central results.
The discussion of the proper regularization scheme is only relevant if
one wants to construct the NZ transition rates from the \textit{T}-matrix expressions.

The exact TCL ME is not equivalent to the approximate Nakajima-Zwanzig-Markov-Pauli ME:
As noted in the discussion of
Eq.\ (\ref{A17.Sexp.20}), the $p=0$ term in this expansion is
$\sum_\mu \tilde\mathcal{R}^{(2\mu)}_\mathrm{reg}$, which we have now
identified as the Nakajima-Zwanzig-Markov-Pauli generator $\tilde\mathcal{G}$.
Using Eqs.\ (\ref{A17.Rregbrack.2}) and (\ref{A17.bbMdef.1}) we can also
write this generator as
\be
\tilde\mathcal{G} \equiv \sum_\mu \tilde\mathcal{R}^{(2\mu)}_\mathrm{reg}
\equiv -i\, [[0]] .
\label{NZ.Gbb0.2}
\ee
The expansion (\ref{A17.Sfull.10}) of the TCL generator $\tilde\mathcal{S}$
contains $\tilde\mathcal{G}$ as the first term but it is followed by an
infinite series of additional terms.

\subsection{The stationary state}
\label{sub.stationary}

Global conservation of probability implies that a stationary solution of any
well-formed Pauli ME exists. Equation (\ref{A4.TCLME.4}) then
shows that the TCL generator $\tilde\mathcal{S}$ must have a right
eigenoperator $\rho_\mathrm{stat}$ to the eigenvalue zero. Due to the $\tP$
projections in $\tilde\mathcal{S}$, this right eigenoperator must be of the form
\be
\rho_\mathrm{stat} = \rho_\mathrm{dot}^\mathrm{stat} \otimes
\rho_\mathrm{leads}^0 ,
\ee
where $\rho_\mathrm{dot}^\mathrm{stat}$ is diagonal.

Applying Eq.\ (\ref{A17.Seqfinal.3}) to $\rho_\mathrm{stat}$, only the $M=0$
term in the sum survives and we obtain $0 = [[0]]\, \rho_\mathrm{stat}$, which
together with Eq.\ (\ref{NZ.Gbb0.2}) implies
\be
\tilde\mathcal{G}\, \rho_\mathrm{stat} = 0 .
\ee
The reverse is also true: If $\tilde\mathcal{G}\, \rho'_\mathrm{stat} = [[0]]\,
\rho'_\mathrm{stat}=0$ then Eq.\ (\ref{A17.Sfull.10}) shows that
$\tilde\mathcal{S}\rho'_\mathrm{stat}=0$.

Thus $\rho_\mathrm{stat}$ is an exact stationary state \emph{if and only if}
$\rho_\mathrm{stat}$ is a right eigenoperator of $\tilde\mathcal{G}$ to the
eigenvalue zero. The exact stationary state
can thus be obtained from the regularized \textit{T}-matrix or
Nakajima-Zwanzig-Markov-Pauli generator $\tilde\mathcal{G}$ alone, in principle.
The formal origin of this result is that all corrections to $\tilde\mathcal{G}$
in $\tilde\mathcal{S}$ contain $\tilde\mathcal{G}$ as the right-most factor,
cf.\ Eq.\ (\ref{A17.Sfull.10}).

There are two caveats, though: (i) The result does not apply to approximations
obtained by truncating the perturbative expansion in $H_\mathrm{hyb}$. It does
work trivially at second order since $\tilde\mathcal{G}^{(2)}=\tilde\mathcal{S}^{(2)}$.
But already at fourth order the TCL Pauli ME for the stationary state reads
\be
-i\,\big( [0] + [0,0,0] - [1][0] \big)\, \rho_\mathrm{stat} = 0 ,
\ee
whereas the Nakajima-Zwanzig-Markov-Pauli ME is
\be
-i\,\big( [0] + [0,0,0] \big)\, \rho_\mathrm{stat} = 0 ,
\ee
which is not equivalent.

(ii) The result does not carry over to time-dependent solutions. Indeed, if
$\rho$ is any eigenoperator of $\tilde\mathcal{S}$ to the eigenvalue
$\lambda$, Eq.\ (\ref{A17.Seqfinal.3}) gives
\be
i\lambda \rho = \sum_{M=0}^\infty (-i\lambda)^M\, [[M]]\, \rho .
\ee
For $\lambda\neq 0$ this does not imply anything for the eigenoperators of
$\tilde\mathcal{G}=-i\,[[0]]$. Conversely, knowing an eigenoperator of
$\tilde\mathcal{G}$ to a non-zero eigenvalue does not help in finding an
eigenoperator of the TCL generator. For the dynamics,
the regularized \textit{T}-matrix or
Nakajima-Zwanzig-Markov-Pauli generator $\tilde\mathcal{G}$ is not sufficient.

\section{Summary and conclusions}
\label{sec.conc}

The dynamics of a quantum dot coupled to electronic leads can be described in the
master-equation formalism. To use this formalism beyond the regime of weak
hybridization between dot and leads, further insight into the structure of
higher-order terms is required. With this motivation, we have derived Pauli master
equations (rate equations) for the probabilites of dot states, to all orders in the
hybridization, and both in time-convolutionless and time-non-local (Nakajima-Zwanzig)
form. Our approach uses a projection superoperator $\tP$ onto product states with
diagonal reduced density matrix. To fourth order, the reduction to the
probabilities has been implemented in Refs.\ \onlinecite{LeW08} and \onlinecite{KGL10}
by explicitly eliminating the off-diagonal components from a Nakajima-Zwanzig-type ME.
Our approach leads to more compact superoperator expressions and is easily
generalized to all orders.

Furtermore, we have presented a superoperator derivation of the
\textit{T}-matrix expression for the Pauli ME and showed that it fails to take
into account the propagation of the density operator from the present time $t$
back to an initial time $t_0$. This answers the question posed in Ref.\
\onlinecite{Tim08} whether it is possible to derive the Pauli master equation
within the \textit{T}-matrix formalism instead of using it \textit{ad hoc} to
calculate the transition rates. The superoperator formalism has allowed us to
establish relationships between the TCL Pauli generator $\tilde\mathcal{S}$, the
NZ generator in the Markov approximation (exact for the stationary state),
$\tilde\mathcal{G}$, and the \textit{T}-matrix generator $\tilde\mathcal{R}$.
The off-diagonal components of these generators are the transition rates in the
respective pictures. Relations between the expansion terms of order $2\mu$,
$\tilde\mathcal{S}^{(2\mu)}$, $\tilde\mathcal{G}^{(2\mu)}$, and
$\tilde\mathcal{R}^{(2\mu)}$, respectively, have been given. In particular, the
expansion terms $\tilde\mathcal{S}^{(2\mu)}$ of the TCL Pauli generator are the
sum of the corresponding terms $\tilde\mathcal{G}^{(2\mu)}$ order by order, plus
corrections, which come from propagating the density operator backward in time
in Eq.\ (\ref{A2.rhoback.2}).
Only at the second (lowest) order the expressions are identical. We have shown
that both the Nakajima-Zwanzig-Markov-Pauli and the TCL Pauli generators
converge in the limit $\eta\to 0^+$, order by order. Here, $\eta$ is the rate
with which the hybridization is switched on. In the NZ case, the absence of
divergences readily emerges from the superoperator expressions, in which the
secular reducible terms are explicitly projected out, whereas for the TCL Pauli
generator it relies on a sweeping cancelation of negative powers of $\eta$.

It is crucial for the derivation that the averages of lead operators satisfy
Wick's
theorem, i.e., that they can be decomposed into averages of pairs. Besides reservoirs
consisting of free fermions as considered here, an analogous derivation should
be possible for free bosons.

As is well known, the \textit{T}-matrix rates diverge for $\eta\to 0^+$.
Specifically, the
term $\tilde\mathcal{R}^{(2\mu)}$ diverges as $1/\eta^{\mu-1}$. The divergence noted for
the fourth-order term by Averin\cite{Ave94} thus becomes even stronger at higher orders.
We have shown that the Nakajima-Zwanzig-Markov-Pauli rates $\tilde\mathcal{G}^{(2\mu)}$
are identical, order by order, to the \textit{T}-matrix rates with proper regularization.
This might lead to an advantage in practical calculations, as the \textit{T}-matrix method
formulated using ordinary operators instead of superoperators is expected to be easier to
implement. This regularization differs from the one proposed by Turek and
Matveev.\cite{TuM02,BKG10,KGL10}

As a consistency check, we have shown that the stationary state obtained from
the Nakajima-Zwanzig-Markov-Pauli ME is the exact one, i.e., is identical to the
stationary solution of the TCL ME. It is quite interesting how this result comes
about: $\tilde\mathcal{S}$ can be written as $\tilde\mathcal{G}$ plus
corrections that all have $\tilde\mathcal{S}$ as the right-most factor again,
cf.\ Eqs.\ (\ref{A17.Seqfinal.3}) and (\ref{NZ.Gbb0.2}). The result does not
carry over to expansions truncated at finite order or to time-dependent
solutions, though.

Another relation between the expansion terms, Eq.\ (\ref{A17.Sfull.10}),
is also important: all correction terms in
$\tilde\mathcal{S}^{(2\mu)}$ beyond $\tilde\mathcal{G}^{(2\mu)}$ have factors
$\tilde\mathcal{G}^{(2\mu')}$ of lower order $\mu'<\mu$ at their right end. Consider the
case that all $\tilde\mathcal{G}^{(2\mu')}$ for $\mu'<\mu_c$ are suppressed but
$\tilde\mathcal{G}^{(2\mu_c)}$ is not. Then all corrections in
$\tilde\mathcal{S}^{(2\mu_c)}$ beyond the Nakajima-Zwanzig-Markov-Pauli term
$\tilde\mathcal{G}^{(2\mu_c)}$ are also small. For example, in the Coulomb-blockade
regime, $\tilde\mathcal{G}^{(2)}$ is suppressed, but $\tilde\mathcal{G}^{(4)}$ is not.
Then the corrections to the cotunneling rates are small in the Coulomb-blockade regime
since they involve sequential-tunneling rates. On the other hand, deep in the
sequential-tunneling regime, all fourth-order terms are small compared to the
sequential-tunneling rates if hybridization is weak. However, close to a
threshold
where some $\tilde\mathcal{G}^{(2\mu')}$ crosses over from small to large, for example at
the Coulomb-blockade threshold, the corrections can be comparable to the NZ rates.

Of course, outside of the perturbative regime there is no \textit{a-priori} reason for
any term to be small. It is in this intermediate coupling regime\cite{RVE10} that we
expect the TCL Pauli ME to all orders to show its power. Since the TCL ME describes
the dynamics exactly, not just the stationary state, and is local in time,
it is promising for resummation schemes addressing for example the dynamical
non-equilibrium Kondo effect.

\acknowledgments

The author would like to thank F. Elste, S. Koller, T. Ludwig, and G. Zar\'and
for helpful discussions and the Asia Pacific Center for Theoretical Physics,
Pohang, for hospitality while part of this work was performed. Financial support
by the Deutsche Forschungsgemeinschaft, in part through Research Unit 1154,
\textit{Towards Molecular Spintronics}, is gratefully acknowledged.

\appendix

\section{Proof of the convergence of certain superoperators for $\eta\to 0^+$}
\label{app.finite}

We want to show that the limit for $\eta\to 0^+$ in Eq.\ (\ref{A17.brackdef.3})
exists and is finite for all non-negative integers
$m_{2\mu-1},m_{2\mu-2},\ldots,m_{2\mu'+1}$, where $\mu>\mu'\ge 0$. It is useful
to prove a more general statement: For $\mu=1,2,\ldots$ the limit for $\eta\to
0^+$ of the superoperator
\ba
\lefteqn{ \tP\,\Lh\, (-\LN+c_{2\mu-1}i\eta)^{-1-m_{2\mu-1}}\,
  \Lh\,\tQ } \nonumber \\
&& {}\times (-\LN+c_{2\mu-2}i\eta)^{-1-m_{2\mu-2}}\,
  \Lh \cdots \nonumber \\
&& {}\times \Lh\,
  (-\LN+c_1 i\eta)^{-1-m_1}\, \Lh\, \tP , \hspace{2.5em}
\label{P1.brk.2}
\ea
where $\tQ$ is inserted at all even-numbered positions, exists and is finite for
all
non-negative integers $m_{2\mu-1},m_{2\mu-2},\ldots,m_1$ and all positive real
numbers $c_{2\mu-1},c_{2\mu-2},\ldots,c_1$. By a finite limit of a superoperator
we mean a finite limit of all its matrix elements.\cite{endnote.super}

By inserting the completeness relation
\be
\bullet = \sum_{ij} |i\rangle\langle i| \bullet |j\rangle\langle j|
  = \sum_{ij} \langle i| \bullet |j\rangle\: |i\rangle\langle j|
\label{P1.compl.1}
\ee
in suitable places, the matrix elements of Eq.\ (\ref{P1.brk.2}) can be
expressed in terms of matrix elements of $\LN$, $\Lh$, and $\tQ$ alone.
The matrix elements of $\LN$ are
\be
\Tr |i\rangle\langle j|\, \LN\, |k\rangle\langle l|
  = \delta_{jk}\delta_{il}\, (E_k-E_l) .
\label{P1.L0mat.3}
\ee
Here, $E_k$ and $E_l$ are eigenenergies of $H_0$, including dot and
lead contributions. The former are discrete and, by assumption,
non-degenerate, whereas the latter have a continuous spectrum. The proposition could
fail if a zero matrix element of $\LN$ occured in Eq.~(\ref{P1.brk.2}).

At this point it is useful to go over to a single-particle description of the
leads.
As noted in Sec.\ \ref{sub.Tm}, the projections $\tP$ in Eq.\ (\ref{P1.brk.2})
introduce equilibrium averages, $\tr_\mathrm{leads}\ldots \rho_\mathrm{leads}^0$, over
lead-electron creation and annihilation operators. These averages are non-zero only
if all lead operators are paired. In the expression (\ref{P1.brk.2}), which is of
order $2\mu$ in $\Lh$, there are $\mu$ such pairs.

Consider a certain inverse superoperator $(-\LN+c_\nu i\eta)^{-1-m_\nu}$ in Eq.\
(\ref{P1.brk.2}). Some of the paired lead operators may straddle its position,
which is numbered by $\nu$. For two paired lead operators that are both to the
right of this position, the superoperator $(-\LN+c_\nu i\eta)^{-1-m_\nu}$ acts
on an operator that is diagonal in the single-electron state associated with the
paired operators. Its energy thus does not appear in the difference $E_k-E_l$ in
Eq.\ (\ref{P1.L0mat.3}). Consequently, only lead-operator pairs that straddle
the position $\nu$ contribute to the energy
difference. Let us denote the number of such pairs by $\zeta_\nu\ge 0$. Then the
difference $E_k-E_l$ has the form $E_n - E_{n'} + \sum_{i=1}^{\zeta_\nu}
\Delta\epsilon_{p_{\nu i}}$, where $E_n$, $E_{n'}$ now denote the energies of the
\emph{dot} many-particle eigenstates $|n)$ and $|n')$, respectively, and the
$\Delta\epsilon_p$, $p=1,2,\ldots,\mu$ are lead single-electron energies. There are $\mu$
such energies, which are independently integrated over from $-\infty$ to $\infty$. The
ordering of the two $\Lh$ insertions where the corresponding lead electron is created and
annihilated determines whether this energy enters with a plus or minus sign in the energy
differences coming from the $\LN$ sandwiched between these two $\Lh$. The
single-particle energy enters with the same sign in all these factors. It
is thus possible to absorb all minus signs into the definitions of
$\Delta\epsilon_p$.

The integrand in the integrals over $\Delta\epsilon_1,\ldots,\Delta\epsilon_\mu$
assumes the general form
\be
\frac{ F\big(\Delta\epsilon_1,\ldots, \Delta\epsilon_\mu\big)}
  {\displaystyle \prod_{\nu=1}^{2\mu-1} \bigg( E_{n_\nu} - E_{n_\nu'} +
  \sum_{i=1}^{\zeta_\nu} \Delta\epsilon_{p_{\nu i}} + c_\nu i\eta
  \bigg)^{1+m_\nu}} ,
\label{P1.genp.4}
\ee
where the function $F$ contains the remaining dependence on the single-electron
energies due to Fermi functions and possibly energy-dependent densities of
states
and tunneling amplitudes. $F$ is assumed to be a real analytic and bounded
function of its arguments. Note that for perfect crystals this does not hold
due to the appearance of van Hove singularities in the density of states.
Any disorder will remove these, though.

We rewrite the expression (\ref{P1.genp.4}) by introducing two sets of
auxilliary variables $x_\nu$ and $\lambda_\nu$,
\begin{widetext}
\ba
\lefteqn{ F\big(\Delta\epsilon_1,\ldots, \Delta\epsilon_\mu\big)\,
  \int dx_1\cdots dx_{2\mu-1} \prod_{\nu=1}^{2\mu-1} \,
  \frac{\ds\delta\left(x_\nu - \sum_{i=1}^{\zeta_\nu} \Delta\epsilon_{p_{\nu
  i}}\right)}{ (E_{n_\nu} - E_{n_\nu'} + x_\nu + c_\nu i\eta)^{1+m_\nu}} }
  \nonumber \\
&& = F\big(\Delta\epsilon_1,\ldots, \Delta\epsilon_\mu\big)\,
  \int dx_1\cdots dx_{2\mu-1} \int \frac{d\lambda_1}{2\pi}\cdots
  \frac{d\lambda_{2\mu-1}}{2\pi}
  \prod_{\nu=1}^{2\mu-1} \,
  \frac{\ds \exp\left(i\lambda_\nu \bigg[x_\nu - \sum_{i=1}^{\zeta_\nu}
  \Delta\epsilon_{p_{\nu i}}\bigg]\right)}{ (E_{n_\nu} - E_{n_\nu'} + x_\nu +
  c_\nu i\eta)^{1+m_\nu}} .
\label{P1.genp.6}
\ea
The integrand as a function of $x_\nu$ has a pole of order $1+m_\nu$
in the negative half plane. Furthermore, it vanishes rapidly for $x_\nu \to
+i\infty$ ($x_\nu \to -i\infty$) if $\lambda_\nu\ge 0$ ($\lambda_\nu\le 0$). The
only possible exception is the case of $m_\nu=0$ and $\lambda_\nu=0$, which we
exclude now and treat separately later. Hence, we can close the integration
contour in the upper (lower) half plane and obtain
\be
\int dx_\nu\: \frac{\ds \exp\left(i\lambda_\nu \bigg[x_\nu -
  \sum_{i=1}^{\zeta_\nu}
  \Delta\epsilon_{p_{\nu i}}\bigg]\right)}{ (E_{n_\nu} - E_{n_\nu'} + x_\nu +
  c_\nu i\eta)^{1+m_\nu}}
  = \left\{\begin{array}{ll}
  0 & \mbox{for } \lambda_\nu\ge 0 \\
  \displaystyle {-\frac{2\pi i}{m_\nu!}}\,
  (i\lambda_\nu)^{m_\nu}
  \exp\left(-i\lambda_\nu\,\bigg[E_{n_\nu} - E_{n_\nu'}
  + \sum_{i=1}^{\zeta_\nu}
  \Delta\epsilon_{p_{\nu i}} + c_\nu i \eta \bigg]\right)
  & \mbox{for } \lambda_\nu\le 0 .
  \end{array}\right.
\label{P1.oneint.5}
\ee
Note that for $m_\nu\ge 1$ the case $\lambda_\nu=0$, which is included in both
lines, is consistent. On the other hand, for $m_\nu=0$ this case was
excluded. We find that Eq.\ (\ref{P1.oneint.5}) shows a step
discontinuity at $\lambda_\nu=0$ for $m_\nu=0$. The result after performing the
integral over $\lambda_\nu$ does not depend on the value at a single point, though.
The expression in Eqs.\ (\ref{P1.genp.4}) and (\ref{P1.genp.6}) now becomes
\ba
\lefteqn{ F\big(\Delta\epsilon_1,\ldots, \Delta\epsilon_\mu\big)\,
  \int_{-\infty}^0 d\lambda_1\,
  \frac{-i^{1+m_1} \lambda_1^{m_1}}{m_1!} \cdots
  \int_{-\infty}^0 d\lambda_{2\mu-1}\,
  \frac{-i^{1+m_{2\mu-1}} \lambda_{2\mu-1}^{m_{2\mu-1}}}{m_{2\mu-1}!} }
  \nonumber \\
&& {}\times  \exp\left(-i \sum_{\nu=1}^{2\mu-1}
  \lambda_\nu\,\bigg[E_{n_\nu} - E_{n_\nu'}
  + \sum_{i=1}^{\zeta_\nu}
  \Delta\epsilon_{p_{\nu i}} + c_\nu i \eta \bigg]\right) \nonumber \\
&& = F\big(\Delta\epsilon_1,\ldots, \Delta\epsilon_\mu\big)\,
  \int_{-\infty}^0 d\lambda_1\,
  \frac{-i^{1+m_1} \lambda_1^{m_1} e^{c_1\eta\lambda_1}}{m_1!}\,
  \exp\left(-i\lambda_1 \left[E_{n_1} - E_{n_1'}\right]\right)
  \cdots \int_{-\infty}^0 d\lambda_{2\mu-1}
  \nonumber \\
&& {}\times \frac{-i^{1+m_{2\mu-1}} \lambda_{2\mu-1}^{m_{2\mu-1}}
  e^{c_{2\mu-1}\eta\lambda_{2\mu-1}}}{m_{2\mu-1}!}\,
  \exp\left(-i\lambda_{2\mu-1} \left[E_{n_{2\mu-1}} -
  E_{n_{2\mu-1}'}\right]\right)
  \exp\left(-i \sum_{p=1}^\mu \left[ \lambda_{\nu_p^-} + \ldots
  + \lambda_{\nu_p^+} \right] \, \Delta\epsilon_p \right) ,\hspace{1em}
\label{P1.genp.8}
\ea
\end{widetext}
where $\nu_p^-$ ($\nu_p^+$) is the first (last) position for which the
single-electron energy $\Delta\epsilon_p$ appears in the energy denominators in
Eq.\ (\ref{P1.genp.4}).

Integrating Eq.\ (\ref{P1.genp.8}) over all $\Delta\epsilon_p$, we obtain
\ba
\lefteqn{ \int_{-\infty}^0 d\lambda_1\,
  \frac{-i^{1+m_1} \lambda_1^{m_1} e^{c_1\eta\lambda_1}}{m_1!}\,
  \exp\left(-i\lambda_1 \left[E_{n_1} - E_{n_1'}\right]\right)
  \cdots } \nonumber \\
&& {}\times \int_{-\infty}^0 d\lambda_{2\mu-1}\,
  \frac{-i^{1+m_{2\mu-1}} \lambda_{2\mu-1}^{m_{2\mu-1}}
  e^{c_{2\mu-1}\eta\lambda_{2\mu-1}}}{m_{2\mu-1}!}  \nonumber \\
&& {}\times \exp\left(-i\lambda_{2\mu-1} \left[E_{n_{2\mu-1}} -
  E_{n_{2\mu-1}'}\right]\right) \nonumber \\
&& {}\times \hat F(\lambda_{\nu_1^-}+\ldots+\lambda_{\nu_1^+}, \ldots,
  \lambda_{\nu_\mu^-}+\ldots+\lambda_{\nu_\mu^+}) \hspace{4em}
\label{P1.genpi.10}
\ea
with the Fourier transform
\ba
\lefteqn{ \hat F(\kappa_1,\ldots,\kappa_\mu)
  := \int d\Delta\epsilon_1\, e^{-i\kappa_1 \Delta\epsilon_1}
  \cdots
  \int d\Delta\epsilon_\mu\, e^{-i\kappa_\mu \Delta\epsilon_\mu} }
  \nonumber \\
&& {}\times F\big(\Delta\epsilon_1,\ldots, \Delta\epsilon_\mu\big) .
  \hspace{14em}
\ea
If there are no lead-operator pairs straddling the position $\nu$, i.e.,
$\zeta_\nu=0$, the variable $\lambda_\nu$ does not occur in $\hat F$ in Eq.\
(\ref{P1.genpi.10}). The integral over $\lambda_\nu$ can then be evaluated and
is proportional to $[c_\nu\eta - i(E_{n_\nu}-E_{n_\nu'})]^{-1-m_\nu}$. If in
addition $E_{n_\nu}-E_{n_\nu'}$ vanishes, we obtain a divergence for $\eta\to
0^+$. But by our assumption of non-degenerate
dot states, $E_{n_\nu}=E_{n_\nu'}$ implies $|n_\nu)=|n_\nu')$. Thus for this
contribution the dot density matrix is diagonal at position $\nu$. Because of
$\zeta_\nu=0$ we can then insert a projection $\tP$ without
changing the result. But there is already a projection $\tQ$ at this
(even-numbered) position and we obtain $\tP\tQ=0$. The divergent term is thus
removed. On the other hand, for $E_{n_\nu}\neq E_{n_\nu'}$ there is no divergence.

It remains to consider the case of at least one lead-operator pair
straddling position $\nu$. Then $\lambda_\nu$ does occur in $\hat F$.
We now consider the properties of the functions $F$ and
$\hat F$. The behavior of $F$ at large $|\Delta\epsilon_p|$ should not affect
the transport and we can therefore assume that $F$ vanishes sufficiently rapidly
and sufficiently smoothly for $\Delta\epsilon_p\to\pm \infty$. We thus assume
that all derivatives $\partial^nF/\partial\Delta\epsilon_p^n$, $n=0,1,2,\ldots$
vanish for $\Delta\epsilon_p\to\pm
\infty$ and that all these derivatives are absolutely integrable in
$\Delta\epsilon_p$ over the real axis. These assumptions require the previously
discussed analyticity property. Under these conditions we have
$\lim_{\kappa_p\to\pm\infty} \kappa_p^n \hat F = 0$
for all $n=0,1,2,\ldots$ and all $p=1,\ldots,\mu$. Thus the Fourier transform
$\hat F$ falls off faster than any power for $\kappa_p\to\pm\infty$ for all $p$.

It follows that $\hat F$ falls off faster than any power for
$\lambda_\nu\to-\infty$. Thus the integral over $\lambda_\nu$ in Eq.\
(\ref{P1.genpi.10}) converges for any $m_\nu$, $E_{n_\nu}-E_{n_\nu'}$,
and $\eta\ge 0$. It thus converges pointwise for $\eta\to 0^+$. The convergence
is also uniform since the integrand in Eq.\ (\ref{P1.genpi.10}) is bounded in absolute
value by the integrand in the expression
\ba
\lefteqn{ \int_{-\infty}^0 d\lambda_1\,
  \frac{\lambda_1^{m_1}}{m_1!}\,
  \cdots \int_{-\infty}^0 d\lambda_{2\mu-1}\,
  \frac{\lambda_{2\mu-1}^{m_{2\mu-1}}}{m_{2\mu-1}!} } \nonumber \\
&& {}\times \big|\hat F(\lambda_{\nu_1^-}+\ldots+\lambda_{\nu_1^+}, \ldots,
  \lambda_{\nu_\mu^-}+\ldots+\lambda_{\nu_\mu^+})\big| \hspace{2em}
\ea
and this integral converges.

In summary, all terms generated by taking the relevant matrix elements of Eq.\
(\ref{P1.brk.2}) and introducing the completeness relation (\ref{P1.compl.1})
remain finite for $\eta\to 0^+$. Since the number of these terms is finite, the
whole quantity remains finite. The convergence is uniform.

\section{Proof of the identity of certain superoperators}
\label{app.indep}

To prove the cancelation of divergences in Sec.\ \ref{sub.cancel} we also need
to show that the superoperators
$[m_{2\mu-1},m_{2\mu-2},\ldots,m_{2\mu'+1}]^{(2\mu,2\mu')}$ defined in Eq.\
(\ref{A17.brackdef.3}) do not depend on the values of the prefactors of $i\eta$,
as long as these are all positive. Therefore, we now prove the following
statement: In the limit $\eta\to 0^+$, the superoperator in Eq.\
(\ref{P1.brk.2}) is independent of $c_{2\mu-1},c_{2\mu-2},\ldots,c_1$ for all
non-negative integers $m_{2\mu-1},m_{2\mu-2},\ldots,m_1$ and all positive real
numbers $c_{2\mu-1},c_{2\mu-2},\ldots,c_1$.

As shown in appendix \ref{app.finite}, this limit is finite. The
derivative of (\ref{P1.brk.2}) with respect to $c_\nu$ is
\ba
\lefteqn{ -i (1+m_\nu) \eta \,
 \tP\,\Lh\, (-\LN+c_{2\mu-1}i\eta)^{-1-m_{2\mu-1}} } \nonumber \\
&& {}\times \Lh\,\tQ \cdots (-\LN+c-\nu i\eta)^{-2-m_\nu}
  \cdots  \Lh\, \tP .\hspace{2em}
\label{A17.fderv.3}
\ea
The derivative and the limit $\eta\to 0^+$ commute because (i) the
expression (\ref{P1.brk.2}) is differentiable with respect to $c_\nu$
for all $\eta>0$, (ii) it converges pointwise for $\eta\to 0^+$ as shown
in appendix \ref{app.finite}, and (iii) its
derivative with respect to $c_\nu$ converges uniformly for $\eta\to 0^+$ (this
is shown by a trivial modification of the proof in appendix \ref{app.finite}
noting that the factor $\eta$ is bounded by unity for $0<\eta\le 1$).

In Eq.\ (\ref{A17.fderv.3}), the part $\tP\cdots\tP$ has a finite limit for $\eta\to 0^+$,
as shown in appendix \ref{app.finite}. Including the extra factor of $\eta$, the limit
vanishes. Consequently, the expression (\ref{P1.brk.2}) is a constant function of
$c_{2\mu-1},c_{2\mu-2},\ldots,c_1$.

\section{Evaluation of prefactors in the superoperator expansion}
\label{app.f}

In this appendix we evaluate the functions $f$ defined in Eq.\ (\ref{A17.f.3}),
which appear as prefactors in the expansion of the TCL generator in powers of
$\eta$. We first consider the case $p=0$, which is more
easily done for the original expression in Eq.\ (\ref{A17.sumint.4}). This
expression does not contain any sum since $n_p=n_0$ is fixed, $n'$
equals zero because of $\mu_p'=\mu_0'=0$, and we obtain
\ba
\lefteqn{ f(n_0;m_{0,2n_0-1},\ldots,m_{0,1}) } \nonumber \\
&& = (2\mu_0-1)^{m_{0,2n_0-1}} (2n_0-2)^{m_{0,2n_0-2}} \cdots 1^{m_{0,1}} .
  \hspace{1.5em}
\ea
Since we are only interested in the case $\Sigma_m=M_0\le p=0$, the only
possibility is $M_0=0$ and thus $m_{0,\nu}=0$ for all $\nu$, giving
$f(n_0;0,\ldots,0) = 1$.

For $p\ge 1$ we evaluate Eq.\ (\ref{A17.f.3}) by iteration. We first perform the
sum over $\pi^Z_0$. The term under the sum is a polynomial in $2\pi^Z_0 \mu_1$
of order $M_0$. The zero-order term in this polynomial vanishes when the sum
is performed due to the factor $(-1)^{\pi^Z_0}$. In particular, for $M_0=0$
this is the only term and the whole expression vanishes,
$f(n_0,n_1,\ldots,n_p;0,\ldots,0,m_{1,2\mu_1-1},\ldots,m_{p,1})=0$.
For $M_0\ge 1$, in all remaining terms of orders $1,\ldots,M_0$ in
$2\pi^Z_0\mu_1$, only the $\pi^Z_0=1$ contribution survives. We thus obtain a
polynomial of order $M_0$ in $2\mu_1$ with the zero-order term missing. It is
thus possible to cancel a factor of $2\mu_1$ with the same factor in the
denominator. What remains is a polynomial in $2\mu_1$ of order $M_0-1\ge 0$.

Now we combine this polynomial with the factors $(2\mu_1-1)^{m_{1,2\mu_1-1}}
(2\mu_1-2)^{m_{1,2\mu_1-2}}\cdots (2\mu_1-2n_1+1)^{m_{1,2\mu_1-2n_1+1}}$
in Eq.\ (\ref{A17.f.3}). These represent a polynomial in $2\mu_1$ of order
$M_1\ge 0$. The product is thus a polynomial of order
$M_0+M_1-1\ge 0$. Using $\mu_1=n_1+\pi^Z_1 \mu_2$, we obtain
polynomials in $2\pi^Z_1 \mu_2$ of order $M_0+M_1-1$. If $M_0+M_1-1=0$ and
$p=1$, $\pi^Z_1=0$ is fixed and we obtain a non-zero result. If $M_0+M_1-1=0$
and $p\ge 2$, we can perform the sum over $\pi^Z_1$. But only the factor
$(-1)^{\pi^Z_1}$ depends on $\pi^Z_1$ and $f$ vanishes.

If $M_0+M_1-1\ge 1$, we necessarily have $\Sigma_m\ge 2$. Then we only have to
consider $p\ge 2$ and there exists a sum over $\pi^Z_1$. As before, the
zero-order term in the polynomial in $2\pi^Z_1 \mu_2$
cancels and the other terms only survive for
$\pi^Z_1=1$. The result is a polynomial in $2\mu_2$ of order
$M_0+M_1-1\ge 1$ with the
zero-order term missing. Canceling a factor $2\mu_2$ with the denominator, we
obtain a polynomial in $2\mu_2$ of order $M_0+M_1-2\ge 0$, which we combine
with the following term to give a polynomial of order $M_0+M_1+M_2-2\ge 0$.
Analogously to the above, if $M_0+M_1+M_2-2\ge 0$ and $p=2$, we obtain a
non-zero result, whereas for $M_0+M_1+M_2-2\ge 0$ and $p\ge 3$ we get $f=0$. If
$M_0+M_1+M_2-2\ge 1$, which requires $\Sigma_m\ge 3$, we iterate these steps.

We obtain $f=0$ if there exists an integer $i < p$ with
\be
M_0+M_1+\ldots+M_i-i = 0 .
\label{A17.cuo.3}
\ee
We obtain $f\neq 0$ if this condition is not satisfied and
\be
M_0+M_1+\ldots+M_p-p \equiv \Sigma_m-p = 0 .
\label{A17.confin.2}
\ee
This implies that $M_0 \ge 1$, $M_0 + M_1 - 1 \ge 1$,
$M_0 + M_1 - 1 + M_2 - 1 \ge 1$, etc.\ and thus
\be
M_0+M_1+\ldots+M_i-i\ge 1
\label{A17.confin.1}
\ee
for all $i<p$. Finally, if
$M_0+M_1+\ldots+M_p-p \equiv \Sigma_m-p < 0$
there must exist an $i<p$ such that condition (\ref{A17.cuo.3}) is satisfied
and we obtain $f=0$.

We draw some conclusions for the case of non-zero $f$ with $\Sigma_m=p$. Since
$M_0+M_1-1+\ldots +M_{p-1}-1\ge 1$ and $M_0+M_1-1+\ldots+M_p-1=0$, we find
$M_p=0$. This implies that $M_0+M_1-1+\ldots+M_{p-1}-1 = 1$. Since further
$M_0+M_1-1+\ldots+M_{p-2}-1 \ge 1$, we conclude that $M_{p-1} \le 1$. By iteration
we find that $M_i \le p-i$.

The next goal is to find the non-zero values of $f$ for all cases with
$\Sigma_m=p$. For $p=0$ we have found $f(n_0;0,\ldots,0)=1$. For $p\ge
1$ we already know that $m_{p,2\mu_p-1}=m_{p,2\mu_p-2}=\ldots=m_{p,1}=0$
is required for a non-zero result. Equation (\ref{A17.f.3}) then assumes the form
\ba
\lefteqn{
f(n_0,n_1,\ldots,n_p;m_{0,2\mu_0-1},\ldots,m_{p-1,2\mu_{p-1}-2n_{p-1}+1},  }
  \nonumber \\
&& 0,\ldots,0) = (-1)^p
  \! \sum_{\pi^Z_0,\pi^Z_1,\ldots,\pi^Z_{p-1}=0}^1
  \frac{\prod_{i=0}^{p-1} \, (-1)^{\pi^Z_i}}
  {2\mu_1 2\mu_2 \cdots 2\mu_p} \nonumber \\
&& {}\times (2\mu_0-1)^{m_{0,2\mu_0-1}} (2\mu_0-2)^{m_{0,2\mu_0-2}}\cdots
  \nonumber \\
&& {}\times (2\mu_0-2n_0+1)^{m_{0,2\mu_0-2n_0+1}} \cdots \nonumber \\
&& {}\times (2\mu_{p-1}-1)^{m_{p-1,2\mu_{p-1}-1}}
  (2\mu_{p-1}-2)^{m_{p-1,2\mu_{p-1}-2}}\cdots \nonumber \\
&& {}\times 1^{m_{p-1,2\mu_{p-1}-2n_{p-1}+1}} . \hspace{8em}
\ea
The factors following the fraction contain exactly $p$ factors of the form
$2\mu_i-\nu = 2(n_i + \pi^Z_in_{i+1} + \pi^Z_i \pi^Z_{i+1} n_{i+2} +
\ldots)-\nu$ with $i\in\{0,\ldots,p-1\}$ and $\nu\in\{1,\ldots,2n_i-1\}$, where
for $m_{i\nu}\ge 2$ we count $m_{i\nu}$ factors. We rewrite this product as
$\prod_{k=0}^{p-1} (2\mu_{i_k} - \nu_k)$,
where we assume, without loss of generality, $0\le i_0\le i_1\le \ldots \le
i_{p-1} \le p-1$ and $\nu_k\le \nu_{k'}$ if $i_k=i_{k'}$ and $k<k'$. Then
the condition $M_0+M_1+\ldots+M_i-i\ge 1$ for $i<p$ implies
$i_k\le k$ for all $k$. Thus we have
\ba
\lefteqn{ f(n_0,n_1,\ldots,n_p;m_{0,2\mu_0-1},\ldots,0) }
  \nonumber \\
&& = (-1)^p \sum_{\pi^Z_0,\pi^Z_1,\ldots,\pi^Z_{p-1}=0}^1
  \prod_{i=0}^{p-1} \, (-1)^{\pi^Z_i}
  \frac{\prod_{k=0}^{p-1} (2\mu_{i_k} - \nu_k)}{\prod_{i=1}^p 2\mu_i} \nonumber
  \\
&& =: \tilde f_p(n_0,n_1,\ldots,n_p;i_0,i_1,\ldots,i_{p-1};
  \nu_0,\nu_1,\ldots,\nu_{p-1}) , \nonumber \\
&&
\ea
where the subscript in $\tilde f_p$ refers to the
number of factors $2\mu_{i_k} - \nu_k$ in the numerator.

By adding and subtracting a constant, we can write
for any $j\in \{0,\ldots,p-1\}$ and any real number $c$,
\ba
\lefteqn{ \tilde f_p(n_0,n_1,\ldots,n_p;i_0,i_1,\ldots,i_{p-1};
  \nu_0,\nu_1,\ldots,\nu_{p-1}) }
  \nonumber \\
&& = \tilde f_p(n_0,\ldots,n_p;i_0,\ldots,i_{p-1}; \nonumber \\
&& \qquad \nu_0,\ldots,\nu_{j-1},c,\nu_{j+1},\ldots,\nu_{p-1}) \nonumber \\
&& {}- (\nu_j-c)\,
  \tilde f_{p-1}(n_0,\ldots,n_p;i_0,\ldots,i_{j-1},i_{j+1},\ldots,i_{p-1};
  \nonumber \\
&& \qquad \nu_0,\ldots,\nu_{j-1},\nu_{j+1},\ldots,\nu_{p-1}) .
\ea
The second term on the right-hand side
contains $\tilde f_{p-1}$, which has $\Sigma_m=p-1$ factors $2\mu_{i_k} -
\nu_k$ in the numerator. But we have shown above that for $\Sigma_m<p$ the
term $f$ vanishes. Thus only the first term remains
and we find that $f=\tilde f_p$ does not depend on $\nu_j$ for any
$j$. Thus we can replace $\nu_j$ by $2(n_{i_j} + n_{i_j+1} + \ldots + n_j)$ (recall
that $i_j\le j$ for all $j$) without changing the value of $f$. We obtain
\ba
\lefteqn{ \tilde f_p(n_0,n_1,\ldots,n_p;i_0,i_1,\ldots,i_{p-1};
  \nu_0,\nu_1,\ldots,\nu_{p-1}) }
  \nonumber \\
&& = (-1)^p \sum_{\pi^Z_0,\pi^Z_1,\ldots,\pi^Z_{p-1}=0}^1
  \prod_{i=0}^{p-1} \, (-1)^{\pi^Z_i} \nonumber \\
&& {}\times \frac{\prod_{k=0}^{p-1} 2(n_{i_k} + \pi^Z_{i_k}\mu_{i_k+1}
  - n_{i_k} - n_{i_k+1} - \ldots - n_k)}{\prod_{i=1}^p 2\mu_i} .
  \nonumber \\[-0.5ex]
&&
\ea
The factor for $k=0$ in the numerator contains $i_k=i_0=0$ and thus
reads $2(n_0 + \pi^Z_0 \mu_1 - n_0)= 2\pi^Z_0 \mu_1$. In the factor for $k=1$ we
have to distinguish the two cases $i_1=0,1$. For $i_1=0$, the
corresponding factor in the numerator reads
$2(n_0 + \pi^Z_0 \mu_1 - n_0 - n_1) = 2(\pi^Z_0n_1 + \pi^Z_0\pi^Z_1 \mu_2-n_1)$.
This factor is multiplied by $\pi^Z_0$ from the $k=0$ factor. Since
$(\pi^Z_i)^2=\pi^Z_i$, we can drop the $\pi^Z_0$ in the $k=1$ factor and write
it as $2(n_1 + \pi^Z_1\mu_2-n_1)=2\pi^Z_1 \mu_2$. If instead $i_1=1$, the $k=1$
factor reads $2(n_1+\pi^Z_1 \mu_2-n_1)=2\pi^Z_1 \mu_2$. We thus obtain the same
result in both cases.

For larger $k$, the factor in the numerator reads
\ba
\lefteqn{ 2(n_{i_k} + \pi^Z_{i_k} n_{i_k+1} + \ldots +
  \pi^Z_{i_k} \pi^Z_{i_k+1} \cdots \pi^Z_k \mu_{k+1} } \nonumber \\
&& {} - n_{i_k} - n_{i_k+1} - \ldots - n_k) .\hspace{8em}
\ea
Since this factor is multiplied by $\pi^Z_0\cdots \pi^Z_{k-1}$ from the
factors for $j<k$, we can drop all $\pi^Z_j$ with $j<k$ and obtain simply
$2\pi^Z_k \mu_{k+1}$.
We finally find
\ba
\lefteqn{ \tilde f_p(n_0,n_1,\ldots,n_p;i_0,i_1,\ldots,i_{p-1};
  \nu_0,\nu_1,\ldots,\nu_{p-1}) }
  \nonumber \\
&& = (-1)^p \sum_{\pi^Z_0,\pi^Z_1,\ldots,\pi^Z_{p-1}=0}^1
  \prod_{i=0}^{p-1} \, (-1)^{\pi^Z_i}
  \frac{\prod_{k=0}^{p-1} 2\pi^Z_k \mu_{k+1}}{\prod_{i=1}^p 2\mu_i}
  \nonumber \\
&& = (-1)^p \, (-1)^p \; = \; 1 .
\ea
We have shown that the coefficients $f$ for $\Sigma_m=p$ vanish if
condition (\ref{A17.cuo.3}) is satisfied and equal unity otherwise.


\begin{thebibliography}{99}

\bibitem{BoW08}L. Bogani and W. Wernsdorfer, Nature Mater.\ \textbf{7}, 179 (2008).

\bibitem{OBL08}E. A. Osorio, T. Bj\o{}rnholm, J.-M. Lehn, M. Ruben, and H. S. J. van
der Zant, J. Phys.: Condens.\ Matter \textbf{20}, 374121 (2008).

\bibitem{AMS10}S. Andergassen, V. Meden, H. Schoeller, J. Splettstoesser, and M. R.
Wegewijs, Nanotechn.\ \textbf{21}, 272001 (2010).

\bibitem{MeW92}Y. Meir and N. S. Wingreen, Phys.\ Rev.\ Lett.\ \textbf{68},
2512 (1992).

\bibitem{WiM94}N. S. Wingreen and Y. Meir, Phys.\ Rev.\ B \textbf{49}, 11040
(1994).

\bibitem{HKH98}M. H. Hettler, J. Kroha, and S. Hershfield, Phys.\ Rev.\ B
\textbf{58}, 5649 (1998).

\bibitem{ScK00}H. Schoeller and J. K\"onig, Phys.\ Rev.\ Lett.\ \textbf{84}, 3686
(2000).

\bibitem{RKW01}A. Rosch, J. Kroha, and P. W\"olfle, Phys.\ Rev.\ Lett.\ \textbf{87},
156802 (2001); A. Rosch, J. Paaske, J. Kroha, and P. W\"olfle, Phys.\ Rev.\ Lett.\
\textbf{90}, 076804 (2003).

\bibitem{MAM04}A. Mitra, I. Aleiner, and A. J. Millis, Phys.\ Rev.\ B
\textbf{69}, 245302 (2004).

\bibitem{PLN05}M. Plihal, D. C. Langreth, and P. Nordlander, Phys.\ Rev.\ B
\textbf{71}, 165321 (2005).

\bibitem{GPM07}R. Gezzi, T. Pruschke, and V. Meden, Phys.\ Rev.\ B \textbf{75}, 045324
(2007).

\bibitem{MiM07}A. Mitra and A. J. Millis, Phys.\ Rev.\ B \textbf{76}, 085342 (2007).

\bibitem{KRS07}T. Korb, F. Reininghaus, H. Schoeller, and J. K\"onig, Phys.\ Rev.\ B
\textbf{76}, 165316 (2007).

\bibitem{JMS07}S. G. Jakobs, V. Meden, and H. Schoeller, Phys.\ Rev.\ Lett.\
\textbf{99}, 150603 (2007).

\bibitem{ElT10}F. Elste and C. Timm, Phys.\ Rev.\ B \textbf{81}, 024421 (2010).

\bibitem{WET08}S. Weiss, J. Eckel, M. Thorwart, and R. Egger, Phys.\ Rev.\ B
\textbf{77}, 195316 (2008).

\bibitem{SMR10}D. Segal, A. J. Millis, and D. R. Reichman, arXiv:1008.5200 (2010).

\bibitem{ScS94}H. Schoeller and G. Sch\"on, Phys.\ Rev.\ B \textbf{50}, 18436
(1994); Physica B \textbf{203}, 423 (1994).

\bibitem{KSS95}J. K\"onig, H. Schoeller, and G. Sch\"on, Europhys.\ Lett.\
\textbf{31}, 31 (1995); Phys.\ Rev.\ Lett.\ \textbf{76}, 1715 (1996).

\bibitem{KSS96}J. K\"onig, J. Schmid, H. Schoeller, and G. Sch\"on,
Phys.\ Rev.\ B \textbf{54}, 16820 (1996).

\bibitem{BrF03}S. Braig and K. Flensberg, Phys.\ Rev.\ B \textbf{68}, 205324 (2003).

\bibitem{BrF04}H. Bruus and K. Flensberg, \textit{Many-body Quantum Theory in
Condensed Matter Physics} (Oxford University Press, Oxford, 2004).

\bibitem{Tim08}C. Timm, Phys.\ Rev.\ B \textbf{77}, 195416 (2008).

\bibitem{LeW08}M. Leijnse and M. R. Wegewijs, Phys.\ Rev.\ B \textbf{78},
235424 (2008).

\bibitem{Sch09}H. Schoeller, Eur.\ Phys.\ J. Special Topics \textbf{168}, 179 (2009).

\bibitem{KGL10}S. Koller, M. Grifoni, M. Leijnse, and M. R. Wegewijs,
arXiv:1008.0347 (2010).

\bibitem{Nak58}S. Nakajima, Prog.\ Theor.\ Phys.\ \textbf{20}, 948 (1958).

\bibitem{Zwa60}R. Zwanzig, J. Chem.\ Phys.\ \textbf{33}, 1338 (1960);
R. Zwanzig, Physica \textbf{30}, 1109 (1964).

\bibitem{ToM76}M. Tokuyama and H. Mori, Prog.\ Theor.\ Phys.\ \textbf{55}, 411
(1976).

\bibitem{STH77}N. Hashitsume, F. Shibata, and M. Shing\={u}, J. Stat.\ Phys.\
\textbf{17}, 155 (1977);
F. Shibata, Y. Takahashi, and N. Hashitsume, \textit{ibid.}\ \textbf{17}, 171
(1977)

\bibitem{Ahn94}D. Ahn, Phys.\ Rev.\ B \textbf{50}, 8310 (1994).

\bibitem{Ake99}H. Akera, Phys.\ Rev.\ B \textbf{60}, 10683 (1999).

\bibitem{GoL04}V. N. Golovach and D. Loss, Phys.\ Rev.\ B \textbf{69}, 245327
(2004).

\bibitem{KOO04}J. Koch, F. von Oppen, Y. Oreg, and E. Sela, Phys.\ Rev.\ B
\textbf{70}, 195107 (2004); J. Koch, F. von Oppen, and A. V. Andreev, Phys.\
Rev.\ B \textbf{74}, 205438 (2006).

\bibitem{JoS06}R. Jorn and T. Seideman, J. Chem.\ Phys.\ \textbf{124}, 084703
(2006).

\bibitem{ElT07}F. Elste and C. Timm, Phys.\ Rev.\ B \textbf{75}, 195341
(2007).

\bibitem{LKO08}M. C. L\"uffe, J. Koch, and F. von Oppen, Phys.\ Rev.\ B \textbf{77},
125306 (2008).

\bibitem{BKG10}G. Begemann, S. Koller, M. Grifoni, and J. Paaske,
Phys. Rev. B \textbf{82}, 045316 (2010).

\bibitem{BrP02}H.-P. Breuer and F. Petruccione, \textit{The Theory of Open
Quantum Systems} (Oxford University Press, Oxford, 2002).

\bibitem{endnote.degen}In order to describe a system with degenerate dot states, we
could introduce a small ad-hoc splitting that is sent to zero at the end of the
calculation, after taking $\eta$ to zero.

\bibitem{Tim09}C. Timm, Phys.\ Rev.\ E \textbf{80}, 021140 (2009).

\bibitem{ScO09}M. G. Schultz and F. von Oppen, Phys.\ Rev.\ B \textbf{80},
033302 (2009).

\bibitem{Buz98}V. Bu\v{z}ek, Phys.\ Rev.\ A \textbf{58}, 1723 (1998).

\bibitem{Kam74}N. G. van Kampen, Physica \textbf{74}, 215 (1974); \textbf{74},
239 (1974).

\bibitem{WaB53}R. K. Wangsness and F. Bloch, Phys.\ Rev.\ \textbf{89}, 728
(1953).

\bibitem{Blo57}F. Bloch, Phys.\ Rev.\ \textbf{105}, 1206 (1957).

\bibitem{Red65}A. G. Redfield, Adv.\ Magn.\ Reson.\ \textbf{1}, 1 (1965).

\bibitem{TuM02}M. Turek and K. A. Matveev, Phys.\ Rev.\ B \textbf{65}, 115332
(2002).

\bibitem{Ave94}D. V. Averin, Physica B \textbf{194--196}, 979 (1994).

\bibitem{RVE10}N. Roch, R. Vincent, F. Elste, W. Harneit, W. Wernsdorfer,
C. Timm, and F. Balestro, Magnetic signature in cotunneling through
$\mathrm{N@C_{60}}$, submitted to Phys.\ Rev.\ Lett.

\bibitem{endnote.super}Matrix elements of superoperators are here understood
with reference to the scalar product of ordinary operators, $\langle A,B\rangle :=
\mathrm{Tr}\, A^\dagger B$. We do not consider problems arising from
infinite-dimensional Fock spaces.

\end{thebibliography}
\end{document}